\documentclass[10pt,journal,compsoc]{IEEEtran}

\pdfoutput=1
\usepackage{amsfonts}
\usepackage{algorithm}
\usepackage{algorithmic}
\usepackage{amsmath}
\usepackage{amssymb}
\usepackage{amsfonts}
\usepackage{bm}
\usepackage{color}
\usepackage{comment}
\usepackage{graphicx}
\usepackage{multirow}
\usepackage{subfigure}
\usepackage{url}
\usepackage{bbding}
\usepackage{pifont}
\usepackage[table]{xcolor}

\ifCLASSOPTIONcompsoc

\else

\fi

\ifCLASSINFOpdf

\else

\fi

\newcommand{\fullname}{\emph{\textbf{I}nformation \textbf{B}ottleneck denoised \textbf{M}ultimedia \textbf{Rec}ommendation~(IBMRec)}}
\newcommand{\shortname}{\emph{IBMRec}}

\begin{document}

\title{Less is More: Information Bottleneck Denoised Multimedia Recommendation}

\author{Yonghui Yang, Le Wu~\IEEEmembership{Member,~IEEE}, Zhuangzhuang He, ZhengWei Wu, Richang Hong,~\IEEEmembership{Senior Member,~IEEE}, Meng Wang,~\IEEEmembership{Fellow,~IEEE}

\IEEEcompsocitemizethanks{

\IEEEcompsocthanksitem  Y.~Yang, L.~Wu, Z.~He, R.~Hong, M.~Wang are with the School of Computer and Information, Hefei University of Technology,
Hefei, Anhui 230009, China.
\protect \\Emails: \{yyh.hfut, lewu.ustc, hyicheng223, hongrc.hfut, \\ eric.mengwang\}@gmail.com. 

\IEEEcompsocthanksitem Z.~Wu is with ByteDance.
\protect \newline Email: zhengwei.wu@bytedance.com.

}

}

\markboth{Journal of xxx }
{Shell \MakeLowercase{\textit{et al.}}: Bare Advanced Demo of IEEEtran.cls for Journals}

\IEEEtitleabstractindextext{
\begin{abstract}
Empowered by semantic-rich content information, multimedia recommendation has emerged as a potent personalized technique. Current endeavors center around harnessing multimedia content to refine item representation or uncovering latent item-item structures based on modality similarity. Despite the effectiveness, we posit that these methods are usually suboptimal due to the introduction of irrelevant multimedia features into recommendation tasks. This stems from the fact that generic multimedia feature extractors, while well-designed for domain-specific tasks, can inadvertently introduce task-irrelevant features, leading to potential misguidance of recommenders.
In this work, we propose a denoised multimedia recommendation paradigm via the \textbf{I}nformation \textbf{B}ottleneck principle~(IB). Specifically, we propose a novel \emph{\textbf{I}nformation \textbf{B}ottleneck denoised \textbf{M}ultimedia \textbf{Rec}ommendation~(IBMRec)} model to tackle the irrelevant feature issue. \shortname~removes task-irrelevant features from both feature and item-item structure perspectives, which are implemented by two-level IB learning modules: feature-level~(FIB) and graph-level~(GIB). In particular, FIB focuses on learning the minimal yet sufficient multimedia features. This is achieved by maximizing the mutual information between multimedia representation and recommendation tasks, while concurrently minimizing it between multimedia representation and pre-trained multimedia features. Furthermore, GIB is designed to learn the robust item-item graph structure, it refines the item-item graph based on preference affinity, then minimizes the mutual information between the original graph and the refined one. Extensive experiments across three benchmarks validate the effectiveness of our proposed model, showcasing high performance, and applicability to various multimedia recommenders.

\end{abstract}}
\maketitle


\IEEEpeerreviewmaketitle

\ifCLASSOPTIONcompsoc
\IEEEraisesectionheading{\section{Introduction}\label{sec:introduction}}
\else

\section{Introduction} \label{sec:intro}
\fi
Learning high-quality user(item) representation is the cornerstone to building modern recommendation systems~\cite{UAI2009BPR, he2016VBPR}. Traditional ID-based recommendation methods focus on learning these representation from the observed interactions, which are widely exploited but usually limited by data sparsity~\cite{UAI2009BPR, he2017NCF, LightGCN}. With the development of deep learning techniques, multimedia-based recommendations are attracting more and more attention, whether on improving recommendation accuracy or alleviating cold-start problems~\cite{he2016VBPR, chen2017ACF, chen2018neural}. In multimedia-based recommendation scenarios, diverse multimedia content is harnessed to enrich and effectively improve personalized services~\cite{he2016VBPR, chen2017ACF, chen2018neural}.

Early works on multimedia recommendation primarily focus on leveraging the extracted multimedia features as side information to enhance item representation, such as VBPR~\cite{he2016VBPR}, and ACF~\cite{chen2017ACF}, which extend Matrix Factorization by incorporating multimedia representation. Recently, some researchers proposed to mine the latent item-item structure based on modality similarity and achieving great success~\cite{zhang2021LATTICE, MICOR}. For example, LATTICE~\cite{zhang2021LATTICE} constructs the item-item graph from their multimodal features, and then injects item affinities into the learning process. In general, multimedia-based recommendations consist of two main steps: extracting multimedia features using pre-trained models and conducting feature-enhanced recommendation learning.

\begin{figure}[t]
    \centering
    \includegraphics[width=88mm]{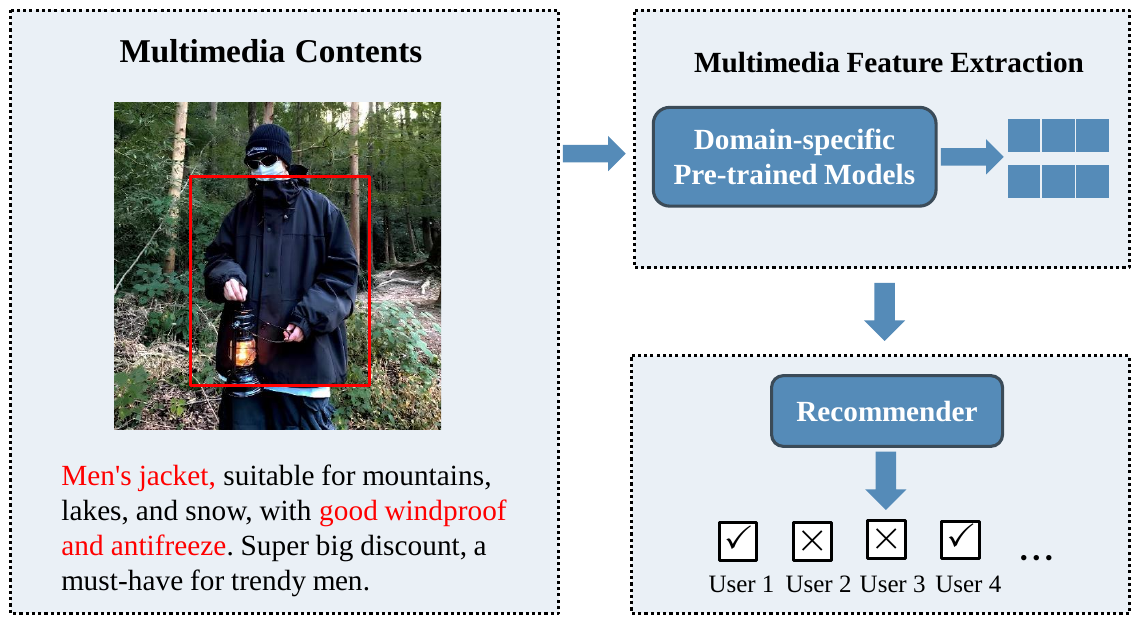}
    \caption{Illustration of irrelevant features on multimedia-based recommendation. Given the multimedia content of an item, where the user's actual focus is highlighted in red, including a black jacket and a performance showcase. Then, the pre-trained models extract multimedia features for recommenders. However, these pre-trained models are well-designed for domain-specific tasks, directly using them will extract irrelevant features for the recommendation task.}
    \label{fig: noise example}
\end{figure}

Despite the effectiveness, we argue that current solutions will introduce irrelevant multimedia features for the recommendation task. As depicted in Fig. \ref{fig: noise example}, we provide a toy example of the irrelevant feature issue on the multimedia-based recommendation. The left side of Fig. \ref{fig: noise example} showcases given multimedia content, with the user's actual focus highlighted in red, including a black jacket and a performance showcase. Typically, domain-specific pre-trained models, such as VGG~\cite{simonyan2014VGG}, BERT~\cite{devlin2018bert}, Sentence Transformers~\cite{reimers2019sentence}, and Hypergraph~\cite{HGSR}, are employed to extract multimedia features. However, these models may not discern the user's true preferences. Utilizing generic multimedia features directly will introduce task-irrelevant information to recommenders~\cite{du2022invariant, huang2023pareto, SGFP}, such as the image background, and redundant text description. To better harness multimedia information, it is crucial to remove the impact of irrelevant features to construct a preference-affinity multimedia recommender.



However, removing the irrelevant multimedia features is non-trivial, as we don't have any prior information to guide what are irrelevant or relevant features for recommenders. Existing works mainly focus on leveraging domain-specific knowledge to elaborate learning processes, such as style-aware fashion recommendation~\cite{liu2017deepstyle} and sentiment-based review recommendation~\cite{lei2016rating}. Nevertheless, these methods heavily rely on domain-specific knowledge, hard to generalize to various recommendation scenarios. Recently, inspired by invariant learning to tackle spurious features~(also irrelevant features), ~\cite{du2022invariant, huang2023pareto} propose multimedia recommendations based on invariant learning, aiming to identify which dimensional features are invariant. However, the extracted features, encompassing both spurious and invariant elements, are indistinguishable from the explicit dimensions. Consequently, removing the impact of irrelevant features from this mixed set remains a challenging task.

\begin{figure*}[t]
\begin{center}
        \includegraphics[width=170mm]{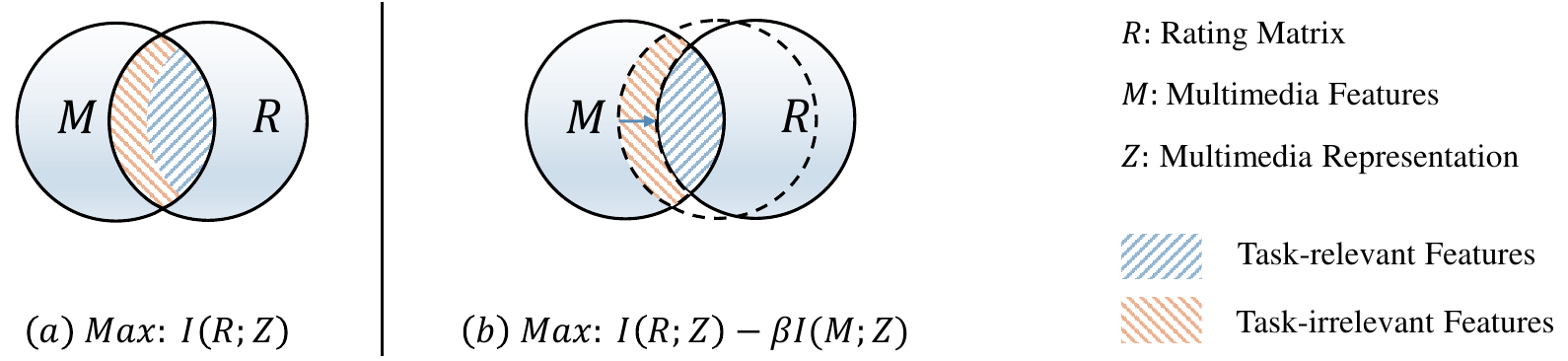}
\end{center}
    \small \caption{Illustration of the optimization objectives. (a) Traditional multimedia recommendation methods only maximize the mutual information between the multimedia representation $\mathbf{Z}$ and rating matrix $\mathbf{R}$, which will introduce irrelevant features to mislead preference learning; (b) Our proposed \shortname~via Information Bottleneck principle, which maximizes the mutual information between the multimedia representation $\mathbf{Z}$ and rating matrix $\mathbf{R}$, meanwhile minimizes it between the multimedia representation $\mathbf{Z}$ and original multimedia feature $\mathbf{M}$. Therefore, \shortname~could learn the minimal sufficient multimedia features for the recommendation task, effectively reducing the impact of irrelevant features.}
    \label{fig: intro_new}
\end{figure*}

In this paper, we propose the \fullname~model to tackle the task-irrelevant features issue. Specifically, we focus on learning the optimal multimedia representation tailored for recommendation tasks. We revisit this problem from an information theory perspective and subsequently propose the \shortname~model via the Information Bottleneck~(IB) principle. As shown in Fig. \ref{fig: intro_new}, we illustrate the optimization objectives of the traditional multimedia recommendation and our proposed \shortname~. 
As depicted in Fig. \ref{fig: intro_new}(a), traditional multimedia recommendation methods only maximize the mutual information between multimedia representation $\mathbf{Z}$ and rating matrix $\mathbf{R}$, denote $I(\mathbf{R}; \mathbf{Z})$. However, simply maximizing $I(\mathbf{R}; \mathbf{Z})$ easily introduces task-irrelevant features to representation learning, as shown in the red part. To this end, we introduce the IB principle to reduce the impact of irrelevant features on the learning process. IB principle describes that an optimal representation should maintain the minimal sufficient information for the downstream task~\cite{tishby2015deep, saxe2019information}. As shown in Fig. \ref{fig: intro_new}(b), our proposed \shortname~maximizes the mutual information of the multimedia representation $\mathbf{Z}$ and rating matrix $\mathbf{R}$, meanwhile minimizes it between the multimedia representation $\mathbf{Z}$ and original multimedia features $\mathbf{M}$. Based on the dual optimization objectives, \shortname~could learn the minimal sufficient multimedia features for the recommendation task, effectively removing the effect of irrelevant features.

Nevertheless, directly optimizing the IB objectives in the multimedia recommendation presents two challenges. First, in traditional IB-based tasks, such as image classification, each sample corresponds to one label. However, in recommendation tasks, the multimedia representation needs to match all users' preferences. Maximizing the mutual information $I(\mathbf{R}; \mathbf{Z})$ is a challenging task due to this multiple matching. Secondly, it's hard to minimize $I(\mathbf{M}; \mathbf{Z})$ because estimating the upper bound of $I(\mathbf{M}; \mathbf{Z})$ is an intractable problem. Although existing methods~\cite{alemi2016VIB, cheng2020CLUB} leverage variational techniques to estimate the upper bound, but heavily rely on the prior assumption, which limits their applications in general scenarios. Third, the irrelevant features lead to an unstable item-item graph structure, which will amplify the irrelevant feature problem through graph convolutions. 

To address the above challenges, we implement \shortname~as follows. Firstly, to overcome the multiple-matching in calculating $I(\mathbf{R}, \mathbf{Z})$, we take userset $U$ and itemset $V$ into formulation, where $I(\mathbf{R}; U, V, \mathbf{Z})=I(\mathbf{R}; U,V)+I(\mathbf{R}; \mathbf{Z}|U,V)$. Then, we derive the lower bounds of $I(\mathbf{R}|U,V)$ and $I(\mathbf{R},\mathbf{Z}|U,V)$, and adopt a two-stage training strategy to optimize the above mutual information objectives. Secondly, for minimizing $I(\mathbf{M}; \mathbf{Z})$, we adopt Hilbert-Schmidt independence criterion~(HSIC)~\cite{ma2020HSIC, wang2021revisiting} to approximate the optimization objective, which doesn't rely on prior experience. Thirdly, we design two-level IB learning modules to adapt graph-based multimedia recommendation. In particular, \shortname~consists of Feature-level IB learning~(FIB) and Graph-level IB learning~(GIB), which can effectively remove the impact of irrelevant features from feature and graph perspectives. The key contributions are summarized as follows:
\begin{itemize}
    \item In this paper, we propose a novel multimedia recommendation method via the Information Bottleneck principle, effectively eliminating the impact of irrelevant features. To the best of our knowledge, we are the first to introduce the IB principle to multimedia recommendation.
    \item Technically, we design a decomposed mutual information maximization solution and introduce the Hilbert-Schmidt independence criterion~(HSIC) to approximate the mutual information minimization.
    \item Given the optimization of mutual information, We implement two-level IB learning modules: feature-level and graph-level IB learning, proficient in removing the impact of irrelevant features from both feature and graph perspectives.
    \item Extensive experiments demonstrate the superiority of the proposed \shortname~, showcasing high performance, and applicability to various multimedia recommenders.
\end{itemize}

\section{Related Works}
\subsection{Multimedia Recommendation}
Representation learning is the key component of modern recommender systems~\cite{UAI2009BPR, he2017NCF, he2016VBPR}. As one of the most popular techniques, Collaborative Filtering learns user and item representation from the observed interactions~\cite{UAI2009BPR, he2017NCF}. However, CF-based methods face the challenge of data sparsity, which is easy to insufficient representation learning. 
Multimedia recommendation extends CF by leveraging information-rich multimedia contents~\cite{he2016VBPR, liu2017deepstyle, chen2017ACF, chen2018neural, deldjoo2020recommender}. In general, these methods use advanced deep learning techniques to extract multimedia features to enhance item representation~\cite{simonyan2014VGG, devlin2018bert, reimers2019sentence}, and learn to match users' content preferences.
For example, VBPR~\cite{he2016VBPR} uses the pre-trained VGGNet to extract visual features to enhance item representation, NARRE~\cite{chen2018neural} uses Word2Vec to extract review features to enhance recommendation precision and explainability. 
Inspired by the great representation power of GNNs~\cite{kipf2016semi}, GNNs have been introduced to recommendation tasks and achieved SOTA performances~\cite{LightGCN, wu2020joint, wu2022graph}. Consequently, a series of graph-based multimedia recommendation methods have been proposed to enhance representation through a combination of graph structure learning and multimedia features~\cite{wei2020GRCN, wei2019MMGCN, zhang2021LATTICE, yi2022MMGCL, wu2020learning, shuai2022review}. LATTICE~\cite{zhang2021LATTICE} is the representative graph-based multimedia recommendation method, which mines latent item-item structure based on their modality similarity, then injects item affinity into preference learning. 

However, few works consider the irrelevant features, a crucial and prevalent problem in multimedia recommendations. As the generic multimedia feature extractors are pre-trained on domain-specific tasks, using the pre-trained extractors inevitably introduces irrelevant features to user preferences~\cite{du2022invariant, huang2023pareto, MICOR, SGFP}. MICOR~\cite{MICOR} proposes a multi-modal contrastive learning framework to force the agreement between each single modality representation and the fused modality representation. Although it can alleviate the irrelevant features from other modalities, it sacrifices the representation ability of all modalities. SGFP~\cite{SGFP} designs a response-based and feature-based distillation loss to transfer knowledge from the generic extractors, but it is still limited by the alignment between multiple modalities. Besides, \cite{du2022invariant, huang2023pareto} propose invariant learning based multimedia recommendations to address the irrelevant feature issue. However, they lie in the assumption that irrelevant features and invariant features are dimensional separate. In practice, the extracted features are mixed, which limits the application of invariant learning on multimedia recommendation. 
Different from the above multimedia recommendation methods, we fully explore irrelevant features from information theory and convert this problem to learn the optimal multimedia representation, which maintains the minimal sufficient features for recommendation tasks.

\subsection{Information Bottleneck}
Information Bottleneck~(IB) is a representation learning principle in the machine learning community based on information theory~\cite{tishby2000information,tishby2015deep,hu2024survey}. For any input data $X$, $Z$ is the hidden representation, and $Y$ is the downstream task label. IB principle describes that a good representation should maintain the minimal sufficient information for the downstream tasks~\cite{tishby2015deep, saxe2019information}: $Max: I(Y; Z) - \beta I(X; Z)$. $I(Y; Z)$ denote the mutual information between the hidden representation $Z$ and label $Y$, $I(X; Z)$ denote the mutual information between the hidden representation $Z$ and input data $X$
two variables, $\beta$ is the coefficient to balance these two parts.

The calculation of the IB has been explored in recent years, especially in deep learning based methods~\cite{alemi2016VIB, saxe2019information, goldfeld2018estimating, cheng2020CLUB}. Precise calculation of mutual information is difficult when dealing with continuous variables. Early works employ variable discrete processing~\cite{tishby2000information} or kernel density estimation~\cite{moon1995estimation} to compute mutual information. 
With the development of deep learning, the more popular solution is approximation estimation, which is widely used in neural networks~\cite{alemi2016VIB, belghazi2018mine, oord2018InfoNCE, cheng2020CLUB}. For estimating the lower bound of mutual information, a series of works are proposed, such as MINE~\cite{belghazi2018mine}, InfoNCE~\cite{oord2018InfoNCE}, and variational-based method~\cite{alemi2016VIB}. 
Besides, few works focus on estimating the upper bound of mutual information, \cite{alemi2016VIB} proposes VIB, a variational approximation to estimate the lower bound of mutual information. Still, it heavily relies on the prior distribution of the latent representation, which limits its application in machine learning. 
Recently, ~\cite{cheng2020CLUB} proposes CLUB to estimate the upper bound of mutual information based on a log-ratio contrastive loss, which is more general for high-dimension learning tasks without any prior. Besides directly optimizing mutual information objectives, researchers propose to use the Hilbert-Schmidt Independence Criterion~(HSIC) to replace mutual information estimation in IB optimization~\cite{ma2020HSIC, wang2021revisiting}. HSIC measures the independence of two variables, which can approximate the mutual information objective.

IB has many applications in machine learning tasks, such as model robustness~\cite{wu2020GIB, wang2021revisiting, zhang2022improving,ding2023robust}, fairness~\cite{gitiaux2021fair, liu2021mitigating, gronowski2023classification}, and explainability~\cite{yu2021recognizing, bang2021explaining, espinosa2022concept, wang2023empower}. In this work, we introduce IB learning to the multimedia recommendation, aiming to reduce the impact of irrelevant features. Considering the difficulty of estimating the upper bound of mutual information, we employ HSIC as the approximation to minimize the mutual information of the pre-trained multimedia feature and its representation.

\section{Preliminaries}

\subsection{Problem Statement}
In a fundamental multimedia recommender, there are two kinds of entities: a userset $U$~($|U|=M$) and an itemset $V$~($|V|=N$), and items associated with $K$-modality multimedia features $\small{\mathbf{M}=\{\mathbf{M}^1,\mathbf{M}^2,...,\mathbf{M}^K\}}$. Considering the recommendation scenarios with implicit feedback, we use matrix {\small{$\mathbf{R} \in \mathbb{R}^{M\times N}$}} to describe user-item interactions, where each element $\mathbf{r}_{ai}=1$ if user $a$ interacted with item $i$, otherwise $\mathbf{r}_{ai}=0$. 
The goal of the multimedia recommendation is to predict users' unknown preferences: $\mathcal{\hat{R}}=f(U, V, \mathbf{Z})=f(U, V, g(\mathbf{M}))$, where $f(\cdot)$ is any recommender learning function and $g(\cdot)$ is multimedia representation learning function. In this work, we focus on learning optimal multimedia representation $\mathbf{Z}$ to facilitate the recommendation task.



\subsection{Hilbert-Schmidt Independence Criterion~(HSIC)}
HSIC (Hilbert-Schmidt Independence Criterion) serves as a statistical measure of dependency~\cite{gretton2005measuring}. This criterion is formulated as the Hilbert-Schmidt norm, assessing the cross-covariance operator between distributions within the Reproducing Kernel Hilbert Space (RKHS). Mathematically, given two variables $X$ and $Y$, $\text{HSIC}(X, Y)$ is defined as follows:
\begin{small}
\begin{equation}
    \begin{aligned}
    HSIC(X, Y) &= ||C_{XY}||_{hs}^2 \\
    &= \mathbb{E}_{X,X',Y,Y'}[K_X(X,X')K_Y(Y,Y')] \\
    &+\mathbb{E}_{X,X'}[K_X(X,X')]\mathbb{E}_{Y,Y'}[K_Y(Y,Y')] \\
    &-2\mathbb{E}_{XY}[\mathbb{E}_{X'}[K_X(X,X')] \mathbb{E}_{Y'}[K_Y(Y,Y')]],
    \end{aligned}
    \label{eq: HSIC}
\end{equation}
\end{small}

\noindent where $K_X$ and $K_Y$ are two kernel functions for variables $X$ and $Y$, $X'$ and $Y'$ are two independent copies of $X$ and $Y$. Given the sampled instances ${(x_i, y_i)}_{i=1}^n$ from the batch training data, the $HSIC(X,Y)$ can be estimated as:
\begin{small}
\begin{equation}
    \hat{HSIC}(X,Y) = (n-1)^{-2}Tr(K_XHK_YH),
\end{equation}
\end{small}

\noindent where $K_X$ and $K_Y$ are used kernel matrices~\cite{gretton2005measuring}, with elements $K_{X_{ij}}=K_X(x_i,x_j)$ and $K_{Y_{ij}}=K_Y(y_i,y_j)$, $H=\mathbf{I}-\frac{1}{n}\mathbf{1}\mathbf{1}^T$ is the centering matrix, and $Tr(\cdot)$ denotes the trace of matrix. In practice, we adopt the widely used radial basis function~(RBF)~\cite{vert2004kernel} as the kernel function:
\begin{equation}
    \label{eq: kernel}
    K(x_i,x_j) = exp(-\frac{||x_i-x_j||^2}{2\sigma^2}),
\end{equation}
\noindent where $\sigma$ is the parameter that controls the sharpness of RBF.

\begin{figure*}[t]
\begin{center} 
        \includegraphics[width=180mm]{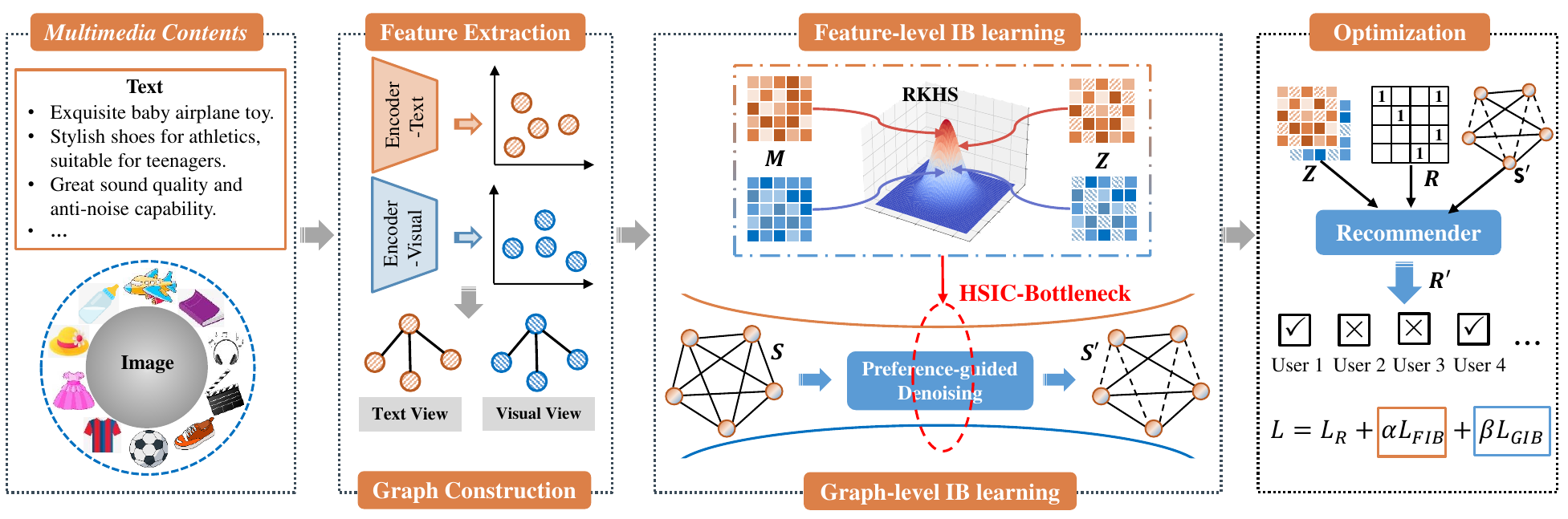}
\end{center}
    \small \caption{Overall framework of our proposed \shortname~, we elaborate two Information Bottleneck learning modules to reduce the impact of irrelevant multimedia features for recommendations tasks: (a) Feature-level IB learning: aiming to remove task-irrelevant multimedia features in representation; (b) Graph-level IB learning: aiming to remove task-irrelevant graph structures to enhance recommendations.}
    \label{fig: framework}
\end{figure*}

\section{Methodology}
In this section, we introduce our proposed \fullname~framework. 
We first present graph-based multimodal representation, followed by two elaborate Information Bottleneck learning modules. Specifically, we design a feature-level Information Bottleneck learning module to remove the redundant features from the pre-trained multimedia features, furthermore, we propose a graph-level Information Bottleneck learning module to filter unstable structures caused by redundant multimedia features. Finally, we show the optimization process of \shortname. The overall architecture is illustrated in Fig. \ref{fig: framework}. 

\subsection{Graph-based Multimodal Representation}
As widely investigated in the previous studies~\cite{LightGCN, zhang2021LATTICE, MICOR}, graph-based methods have shown great potential in multimedia recommendations. Given the user-item interactions and extracted multimodal features, we conduct collaborative user-item graph learning and semantic-aware item-item graph learning to obtain informative multimodal representation.


\subsubsection{Collaborative User-Item Graph Learning}
Given the observed user-item interactions $\mathbf{R}$, we first construct a collaborative user-item graph $\mathcal{G}=\{U \cup V, \mathbf{A}\}$, where $U \cup V$ denotes the set of nodes, and $\mathbf{A}$ is the adjacent matrix defined as follows:
\begin{small}
\begin{flalign}\label{eq:adj_matrix}
\mathbf{A}=\left[\begin{array}{cc}
\mathbf{0}^{M\times M} & \mathbf{R}\\
\mathbf{R}^T & \mathbf{0}^{N\times N}
\end{array}\right].
\end{flalign}
\end{small}

\noindent Let matrices $\mathbf{P^0} \in \mathbb{R}^{M \times d_1}$ and $\mathbf{Q^0} \in \mathbb{R}^{N \times d_1}$ denote the initialized user and item latent embeddings, and matrices $\mathbf{C}^0$ and $\mathbf{Z}^0$ denote the initialized user and item multimedia embeddings, where $\mathbf{Z}^0=g(\mathbf{M})$. We initialize user and item embedding matrices as follows:
\begin{small}
\begin{flalign}
        \mathbf{X}^0 = [\mathbf{P}^0, \mathbf{C}^0],
        \mathbf{Y}^0 = [\mathbf{Q}^0, \mathbf{Z}^0].
\end{flalign}
\end{small}
 

\noindent Then, we update user and item embeddings through multiple graph convolutions:
\begin{small}
\begin{flalign}
    \left[\begin{array}{cc}
    \mathbf{X}^{l+1} \\
    \mathbf{Y}^{l+1}
    \end{array}\right] = \mathbf{D}_A^{-\frac{1}{2}}\mathbf{A}\mathbf{D}_A^{-\frac{1}{2}}     \left[\begin{array}{cc}
    \mathbf{X}^l \\
    \mathbf{Y}^l
    \end{array}\right],
\end{flalign}
\end{small}

\noindent where $\mathbf{D}_A$ is the degree matrix of the adjacent matrix $\mathbf{A}$. After stacking $L$ convolution layers, we have $L+1$ user~(item) embedding matrices:$\{\mathbf{X}^0, ..., \mathbf{X}^L\}$ and $\{\mathbf{Y}^0, ..., \mathbf{Y}^L\}$, then we obtain user embeddings $\mathbf{X}$ and item embeddings $\mathbf{Y}$ after graph convolutions:
\begin{small}
    \begin{equation}
    \label{eq: readout}
    \mathbf{X}=\frac{1}{L+1}\sum_{l=0}^L\mathbf{X}^l,
    \mathbf{Y}=\frac{1}{L+1}\sum_{l=0}^L\mathbf{Y}^l.
\end{equation}
\end{small}

\subsubsection{Semantic-aware Item-Item Graph Learning}
To further enhance item representation with semantic correlations, we follow~\cite{zhang2021LATTICE, MICOR} and construct the semantic-aware item-item graph based on items' multimodal features. Specifically, we employ the kNN technique to construct the semantic matrix $\mathbf{S}^m$ for $m^{th}$ modality. Each element $\mathbf{S}^m_{ij}$ is firstly computed by cosine similarity:
\begin{small}
    \begin{equation}
        \label{cos similarity}
        \mathbf{S}^m_{ij} = \frac{(\mathbf{Z}^m_i)^T\mathbf{Z}^m_j}{||\mathbf{Z}^m_i|| \cdot ||\mathbf{Z}^m_j||},
    \end{equation}
\end{small}

\noindent Then, we use kNN to preserve the Top-K similar semantic neighbors for each item:
\begin{small}
\begin{flalign}\label{eq:topk_sparse}
\mathbf{S}^m_{ij} = 
\begin{cases}
\mathbf{S}^m_{ij}, \quad \mathbf{S}^m_{ij}\in {TopK(\mathbf{S}^m_{i})} \\
0, \quad \mathbf{S}^m_{ij}\notin {TopK(\mathbf{S}^m_{i})},
\end{cases}
\end{flalign}
\end{small}

After obtaining the sparsified item-item correlation matrix $\mathbf{S}^m$ in each modality, we fuse them to obtain the final correlation matrix $\mathbf{S}$:
\begin{small}
    \begin{equation}
        \mathbf{S} = \sum_{m}\mathbf{W}_m\mathbf{S}^m,
    \end{equation}
\end{small}

\noindent where $\mathbf{W}_m$ is the learnable weight matrix in $m^{th}$ modality. Given the initialized item embedding matrix ${\mathbf{Y'}^0}=\mathbf{Y}^0$, we update it as follows:
\begin{small}
\begin{flalign}
{\mathbf{Y'}^{l}}=\mathbf{D}_S^{-\frac{1}{2}}\mathbf{S}\mathbf{D}_S^{-\frac{1}{2}} {\mathbf{Y'}}^{l-1}, 
\end{flalign}
\end{small}

        

\noindent where $\mathbf{D}_S$ is the degree matrix of the correlation matrix $\mathbf{S}$. We define $\mathbf{Y'}=\mathbf{Y'}^L$ as the learned item embeddings after $L$ layer item-item convolution operations. Finally, we fuse the learned item embedding matrices from the user-item graph and item-item graph as the final item representation:
\begin{small}
\begin{equation}
    \mathbf{Y} = \mathbf{Y} + \frac{{\mathbf{Y'}}}{||{\mathbf{Y}}||}.
\end{equation}
\end{small}

\subsubsection{Limitations of Current Representation Learning}
Based on the above descriptions of collaborative user-item graph learning and semantic item-item graph learning, we can summarize the current graph-based multimodal learning process as follows:
\begin{small}
    \begin{equation} 
        \label{eq: gnn_learning}
        \mathbf{X,Y}=GNN(\mathbf{A},\mathbf{S},\mathbf{M}).
    \end{equation}
\end{small}
\noindent In summary, the current learning paradigm mainly focuses on improving item representation quality from explicit feature enhancement and implicit item-item affinity learning. Nevertheless, this solution is still sub-optimal due to the inevitable introduction of irrelevant multimedia features. The reason is that the input multimedia features are extracted from the domain-specific pre-trained models, which are not considered designed for recommendation tasks. 

To this end, we elaborate on two-level Information Bottleneck learning modules from feature and graph perspectives, aiming to reduce the noise of irrelevant multimedia features for recommendation tasks. Next, we introduce each Information Bottleneck learning module, respectively.

\subsection{Feature-level Information Bottleneck Learning}
In this part, we introduce the Feature-level Information Bottleneck~(FIB) Learning module to remove the redundant part contained in the input multimedia features. Specifically, given the input multimedia features $\mathbf{M}$, FIB requires multimedia representation $\mathbf{Z}$ to maximize the mutual information to rating matrix $\mathbf{R}$ and meanwhile minimize it to input multimedia features $\mathbf{M}$. The overall learning objective is described as:
\begin{small}
    \begin{equation}
        Max: I(\mathbf{R}; \mathbf{Z}) - \alpha I(\mathbf{Z}; \mathbf{M}).
    \end{equation}
\end{small}

\noindent However, directly calculating the mutual information is challenging due to the following: (1) Multiple matching: the multimedia representation needs to match all users' preferences; (2) Lacking prior knowledge of $p(\mathbf{Z})$ lead to inaccurate estimation of variational bounds of mutual information. Focusing on these two challenges, we introduce FIB learning in the following.

\subsubsection{Maximization of $I(\mathbf{R}; \mathbf{Z})$}
Here, we consider the characteristics of recommendation tasks that a
good multimedia representation serves all user-item preferences. 
Therefore, we use $I(R; U, V, \mathbf{Z})$ to instead $I(R, \mathbf{Z})$. According to the Markov chain property of mutual information, we decompose $I(R; U, V, \mathbf{Z})$ as follows:
\begin{small}
\begin{flalign}
    \label{eq: Max MI}
    I(R;U,V,\mathbf{Z}) &= I(R;U,V) + I(R;\mathbf{Z}|U,V).
\end{flalign}
\end{small}

\noindent For the first term of Eq.\eqref{eq: Max MI}, we derivate the lower bound of $I(R;U,V)$ as follows:
\small
\begin{equation}
    \begin{aligned}
         I(\mathbf{R}; U,V) & \overset{(a)}{=} H(\mathbf{R})-H(\mathbf{R}|U,V)\\
    & \overset{(b)}{\geq} \sum_{a \in U}\sum_{i \in V} \sum_{r\in R} p(r,a,i)log(p(r|a,i)) \\ 
    & \overset{(c)}{\geq} 
    \sum_{(a,i,j) \in \mathcal{D}}log(p(r=1|a,i))+log(p(r=0|a,j)) \\
    & \overset{(d)}{=} \sum_{(a,i,j) \in \mathcal{D}}log(\sigma(\hat{r}_{ai})) - log(\sigma(\hat{r}_{aj})), \\
    & \overset{(e)}{\geq} \sum_{(a,i,j) \in \mathcal{D}}log(\sigma(\hat{r}_{ai}-\hat{r}_{aj})),   
    \end{aligned}
\end{equation}


\noindent where $\hat{r}_{ai}=p(r_{ai}=1|a,i)$ denotes the propensity score that user $a$ will interacted with item $i$, $\sigma(\cdot)$ denotes the sigmoid activation, and $\mathcal{D}=\{(a,i,j)|r_{ai}=1 \!\wedge\! r_{aj}=0\}$ is training data.
From the above derivation, we obtain a lower bound of $I(R;U,V)$. Given embedding-based recommender systems~(i.e., VBPR~\cite{he2016VBPR}, LATTICE~\cite{zhang2021LATTICE}), we minimize the common BPR ranking loss~\cite{UAI2009BPR} instead of maximizing the mutual information of $I(\mathbf{R};U,V)$.

For the second term of Eq.\eqref{eq: Max MI}, given batch training samples $\mathcal{B}=\{a,i,\mathbf{z}_i,r\}$, we have the lower bound of $I(\mathbf{R}; \mathbf{Z}|U,V)$ as follow:
\begin{small}
    \begin{equation}
    \begin{aligned} \label{eq: second term}
    I(\mathbf{R};\mathbf{Z}|U,V) &\geq log N - \mathcal{L}_N \\
    &=log N + \mathbb{E}_r log \frac{p(r_{ai}|\mathbf{z}_i, a, i)/p(r_{ai}|a,i)}{\sum_{j \in \mathcal{B}_v}p(r_{ai}|\mathbf{z}_j, a, i)/p(r_{ai}|a,i)}
    \end{aligned}
    \end{equation}
\end{small}

\noindent where $N$ denote the batch size, and $\mathcal{B}_v$ denote batch items. We proof Eq.\eqref{eq: second term} as follows:
\begin{small}
    \begin{equation}
    \begin{aligned}
    \mathcal{L}_N &= -\mathbb{E}_r log \frac{p(r_{ai}|\mathbf{z}_i, a, i)/p(r_{ai}|a,i)}{\sum_{j \in \mathcal{B}_v} p(r_{ai}|\mathbf{z}_j,a,i)/p(r_{ai}|a,i)} \\
    &= \mathbb{E}_r log [1+\frac{p(r_{ai}|a,i)}{p(r_{ai}|\mathbf{z}_i,a,i)}\sum_{j!=i}\frac{p(r_{ai}|\mathbf{z}_j,a,i)}{p(r_{ai}|a,i)}] \\
    &\approx \mathbb{E}_r log[1+\frac{p(r_{ai}|a,i)}{p(r_{ai}|\mathbf{z}_i,a,i)}(N-1)\mathbb{E}\frac{p(r_{ai}|\mathbf{z}_j,a,i)}{p(r_{ai}|a,i)}] \\
    &=\mathbb{E}_r log[1+\frac{p(r_{ai}|a,i)}{p(r_{ai}|\mathbf{z}_i,a,i)}(N-1)] \\
    &\geq \mathbb{E}_r log\frac{p(r_{ai}|a,i)}{p(r_{ai}|\mathbf{z}_i,a,i)} + log N \\
    &= -I(\mathbf{R};\mathbf{Z}|U,V)+logN.
    \end{aligned}
    \end{equation}
\end{small}

Thus, we minimize $\mathcal{L}_N$ to implement the maximization of the mutual information of $I(\mathbf{R};\mathbf{Z}|U,V)$. The key challenge is to estimate the density ratio of $p(r_{ai}|\mathbf{z}_i,a,i)/p(r_{ai}|a,i)$. Here, we propose conditional noise contrastive estimation instead of the common noise contrastive estimation~\cite{oord2018InfoNCE}. Following the basis assumption of multimodal recommendation~\cite{he2016VBPR, chen2017ACF}, the user's preference depends on two views: collaborative preference view and multimedia preference view. For each $(a,i)$ pair, we use $f(a,i)$ and $g(a,\mathbf{z}_i)$ to denote the collaborative and multimedia preference score, the overall score $p(r_{ai}=1|a,i,\mathbf{z}_i)=f(a,i)+g(\mathbf{c}_a,\mathbf{z}_i)$. Then, the density ratio $p(r_{ai}|\mathbf{z}_i,a,i)/p(r_{ai}|a,i)$ can be decomposed as:
\begin{small}
    \begin{equation}
    \begin{aligned}
    \frac{p(r_{ai}|\mathbf{z}_i,a,i)}{p(r_{ai}|a,i)} &= \frac{p(r_{ai}|\mathbf{z}_i,a,i)}{\sum_{j \in \mathcal{B}_v}\frac{1}{N}p(r_{ai}|\mathbf{z}_j,a,i)} \\
    &= \frac{f(a,i)+g(\mathbf{c}_a,\mathbf{z}_i)}{f(a,i)+\mathbb{E}g(\mathbf{c}_a,\mathbf{z}_j)}.
    \end{aligned}
    \end{equation}
\end{small}

\noindent Given the trained model $f(,)$, the above objective equals to estimate $g(\mathbf{c}_a,\mathbf{z}_i)/{\mathbb{E}g(\mathbf{c}_a, \mathbf{z}_j)}$. Next, we use $exp({<\mathbf{c}_a, \mathbf{z}_i>}/{\tau})$ to approximate the density ratio ${p(r_{ai}|\mathbf{z}_i,a,i)}/{p(r_{ai}|a,i)}$, where $<,>$ denotes cosine similarity function and $\tau$ denotes the temperature parameter. Based on the above analysis, we obtain the maximization optimization objective of $I(\mathbf{R};\mathbf{Z}|U,V)$:
\begin{small}
    \begin{equation}
   -\frac{1}{N}\sum_{(a,i) \in \mathcal{B}} log\frac{exp(<\mathbf{c}_a,\mathbf{z}_i>/\tau)}{\sum_{j \in \mathcal{B}_v}exp(<\mathbf{c}_a,\mathbf{z}_j>/\tau)}.
    \end{equation}
\end{small}

\noindent In practice, we adopt two-stage training to maximize $I(\mathbf{R};U,V)$
and $I(\mathbf{R};\mathbf{Z}|U,V)$. In the first phase, we update all parameters to satisfy the recommendation task. In the second phase, we only update the parameters of $g(\cdot,\cdot)$, which can better distinguish the user's preferences for different multimedia contents.

\subsubsection{Minimization of $I(\mathbf{Z}, \mathbf{M})$}
In this part, we introduce how to minimize $I(\mathbf{Z}, \mathbf{M})$, aiming to remove the redundant multimedia features for recommendation tasks. 
Specifically, given the extracted generic multimedia features $\mathbf{M}={\mathbf{M}^1,\mathbf{M}^2,...,\mathbf{M}^K}$, we first use $K$ MLP layers to obtain their representation $\mathbf{Z}$:
\begin{small}
\begin{equation}
\begin{aligned}
    \label{eq: multimedia representation}
    \mathbf{Z}&=[\mathbf{Z}^1, \mathbf{Z}^2, ..., \mathbf{Z}^K] \\
    &=[MLP_1(\mathbf{M}^1), MLP_2(\mathbf{M}^2,..., MLP_K(\mathbf{M}^K))].
\end{aligned}
\end{equation}
\end{small}

As $\mathbf{Z}$ is a high-dimensional variable with infinite support, calculating the mutual information $I(\mathbf{Z}, \mathbf{M})$ is intractable. Therefore, the current solution is to minimize the upper bound of $I(\mathbf{Z}, \mathbf{M})$ instead of directly minimizing $I(\mathbf{Z}, \mathbf{M})$. However, estimating the upper bound of the mutual information is a difficult problem, because the variational methods heavily rely on the prior distribution $p(\mathbf{Z})$~\cite{alemi2016VIB} and $p(\mathbf{Z},\mathbf{M})$~\cite{cheng2020CLUB}. Besides, the sampling quality limits the estimation of the upper bound~\cite{alemi2016VIB,cheng2020CLUB}. Inspired by~\cite{ma2020HSIC}, here we introduce Hilbert-Schmidt Independence Criterion~(HSIC) to approximate the mutual information minimization. We use HSIC to constrain each modality multimedia feature learning separately, then we have feature-level information bottleneck regularization:
 \begin{small}
\begin{equation}
    \label{eq: FIB_Loss}
    \mathcal{L}_{FIB}=\hat{HSIC}(Z,M) = \sum_{k=1}^K \hat{HSIC}(Z^k, M^k).
\end{equation}
\end{small}

\subsection{Graph-level Information Bottleneck Learning}
Besides directly removing the irrelevant multimedia features via feature-level information bottleneck learning, we further design a graph-level information bottleneck learning module against the unstable item-item correlation graph. As we introduced in semantic-aware item-item graph learning, the construction of the item-item correlation graph relies on the similarity of the extracted multimedia features. However, this item-item affinity is unstable, even amplifying the irrelevant multimedia issue. Therefore, we design graph-level information bottleneck learning~(GIB), aiming to learn a robust item-item graph for multimedia recommendations. Specifically, GIB consists of two components: preference-guided structure learning and GIB-based structure optimization.


\subsubsection{Preference-guided Structure Learning}
Instead of constructing the item-item correlation graph by directly using multimedia features, here we inject collaborative information to refine the item-item graph structure. Given the adjacent matrix $\mathbf{S}$ of the item-item graph, we aim to learn the masked graph structure $\mathbf{S'}$:
\begin{small}
\begin{equation}
    \mathbf{S'} = \{\mathbf{s}_{ij} \odot \rho_{ij}\},
\end{equation}
\end{small}

\noindent in which $\rho_{ij} \sim Bern(w_{ij})$ denotes that each edge is dropped with the probability $1-w_{ij}$. We combine the denoised multimedia representation and collaborative representation to refine the graph structure. For each item pair $<i,j>$, We adopt multi-layer perceptrons~(MLPs) to learn the distribution parameter $w_{ij}$:
\begin{small}
\begin{equation}
    w_{ij} = MLP(f(\mathbf{z}_i,\mathbf{q}_i), f(\mathbf{z}_j,\mathbf{q}_j),
\end{equation}
\end{small}

\noindent where $\mathbf{z}$ and $\mathbf{q}$ denote the item's denoised multimedia representation and collaborative representation, respectively. $f(,)$ is the fusion function of multimedia and collaborative signals. In practice, we use the concatenate function and achieve good performances. 

\subsubsection{GIB-based Structure Optimization}
Although refining the item-item graph structure with both collaborative and multimedia representation, how to optimize the above learning process is still challenging due to a lack of supervision. Here, we introduce our optimization solution based on GIB: \textit{learning the minimal sufficient graph structure for the recommendation task}. The overall optimization objective of GIB-based recommendation is defined as follows: 
\begin{small}
\begin{equation}
    Max: I(\mathbf{R};U,V,\mathbf{S}')-\beta I(\mathbf{S}';\mathbf{S}),
\end{equation}
\end{small}

\noindent where $\beta$ is the balance parameter, $ I(\mathbf{R};U,V,\mathbf{S}')$ aims to learn the sufficient graph structure for recommendation tasks, and the $ I(\mathbf{S}';\mathbf{S})$ aims to learn the minimal structure information from the input item-item graph. The first term of GIB is easily optimized, we can feed the preference-guided graph structure $S'$ into the representation learning process~(Eq.(13)), and then $Max: I(\mathbf{R}; U, V, \mathbf{S}')$ just equals $Max: I(\mathbf{R}; U,V)$ ~(Eq.(16)).
The challenge of GIB optimization lies in the second term: $I(\mathbf{S}';\mathbf{S})$. The reason is that the discrete graph structures hard to compute the mutual information. Therefore, we relax the optimization of minimizing the mutual information between the input graph and the learned graph as follows:
\begin{small}
\begin{equation}
    Min: I(\mathbf{S}';\mathbf{S}) = \sum_{i=0}^{N-1} I({\mathbf{Y'}}; {\mathbf{Y}}),
\end{equation}
\end{small}

\noindent where ${\mathbf{Y'}}$ and ${\mathbf{Y}}$ denote the item representation given $S'$ and $S$ as GNN inputs, respectively:
\begin{small}
\begin{equation}
    {\mathbf{Y'}}=GNN(\mathbf{A},\mathbf{S}',\mathbf{M}), {\mathbf{Y}}=GNN(\mathbf{A},\mathbf{S},\mathbf{M}),
\end{equation}
\end{small}
\noindent Thus, we decompose graph-level IB learning into node-level IB learning. We approximate the minimization of $I({\mathbf{Y'}}; {\mathbf{Y}})$ by minimizing the $HSIC({\mathbf{Y'}}; {\mathbf{Y}})$. Then, we have graph-level information bottleneck regularization:
\begin{small}
    \begin{equation}
        \mathcal{L}_{GIB}=\hat{HSIC}(\mathbf{S}';\mathbf{S}')\approx \hat{HSIC}(\mathbf{Y}';\mathbf{Y}).
    \end{equation}
\end{small}


\subsection{Model Optimization}
Given the learned user and item embedding matrices $\mathbf{X}$ and $\mathbf{Y}'$~(learned from the preference-guided graph $S'$), we compute the preference score $\hat{r}_{ai}=<\mathbf{x}_a, \mathbf{y}'_i>$, where $<,>$ is the inner product. We select popular BPR ranking loss~\cite{UAI2009BPR} to optimize the maximization of $I(\mathbf{R};U,V)$:
\begin{equation}
\label{eq: loss_rec}
    \mathop{\arg\min}\limits_{\Theta_1} \mathcal{L}_{rec} = \sum_{a=0}^{M-1}\sum\limits_{(i,j)\in D_a }-log\sigma(\hat{r}_{ai}-\hat{r}_{aj}) + \lambda ||\Theta_1||^2,
\end{equation}

\noindent  where $\Theta_1=[\mathbf{P},\mathbf{Q},\mathbf{C},\mathbf{Z}]$,
$\sigma(\cdot)$ is the sigmoid activation function, $\lambda$ is the regularization coefficient.
{\small$D_a=\{(i,j)|i\in R_a\!\wedge\!j\not\in R_a\}$} denotes the pairwise training data for user $a$. {\small$R_a$} represents the item set that user $a$ has interacted. According to Eq.(\ref{eq: loss_rec}), we realize the maximization of $I(\mathbf{R};U,V)$. Combining the feature-level and graph-level HSIC-based bottleneck regularization, we have the overall optimization objective:
\begin{small}
\begin{equation}
\mathop{\arg\min}\limits_{\Theta_1}\mathcal{L}_1=\mathcal{L}_{rec}+\alpha \hat{HSIC}(\mathbf{Z},\mathbf{M}) + \beta \hat{HSIC}(\widetilde{\mathbf{Y'}}; \widetilde{\mathbf{Y}}), 
        \label{eq: all_loss}
\end{equation}
\end{small}

\noindent where $\alpha$ is the FIB loss coefficient, and $\beta$ is the GIB loss coefficient. These two parameters are used to balance the recommendation task and the reduction of irrelevant features. Besides, we optimize the maximization objective of $I(\mathbf{R};\mathbf{Z}|U,V)$:
\begin{small}
    \begin{equation}
   \mathop{\arg\min}\limits_{\Theta_2} \mathcal{L}_2 = -\frac{1}{N}\sum_{(a,i) \in \mathcal{B}} log\frac{exp(<\mathbf{c}_a,\mathbf{z}_i>/\tau)}{\sum_{j \in \mathcal{B}_v}exp(<\mathbf{c}_a,\mathbf{z}_j>/\tau)},
    \end{equation}
\end{small}

\noindent where $\Theta_2=[\mathbf{C},\mathbf{Z}]$.
We optimize $\mathcal{L}_1$ and $\mathcal{L}_2$ iteratively, the first term aims to reduce the irrelevant multimedia features in the recommendation task, and the second aims to fully exploit the multimedia feature to refine user's preferences. The overall optimization process is illustrated in Algorithm 1.

\begin{algorithm}[t]
\renewcommand{\algorithmicrequire}{\textbf{Input:}}
\renewcommand\algorithmicensure {\textbf{Output:}}
\caption{\small{The Algorithm of \shortname}}\label{alg: ibrec}
    \begin{algorithmic}[1]
    \REQUIRE user-item interactions $\mathbf{R}$, extracted multimedia features $\mathbf{M}={M^1,...,M^K}$;  ~~\
    \ENSURE Parameters $\Theta_1=[\mathbf{P}, \mathbf{Q}, \mathbf{C}, \mathbf{Z}]$~~\\
    \STATE Construct the item-item correlation graph $\mathbf{S}$ based on modality similarity;
    \WHILE{not converged}
    \STATE Sample a mini-batch training data;
    \STATE Compute preference-guided item-item graph $\mathbf{S}'$~(Eq.(23-24));
    \STATE Feed $\mathbf{S}'$ as input to compute user and item representation~(Eq.(\ref{eq: gnn_learning}));
    \STATE Compute the recommendation task loss $\mathcal{L}_{rec}$~(Eq. \eqref{eq: loss_rec});
    \STATE Compute feature-level information bottleneck loss $\hat{HSIC(\mathbf{Z},\mathbf{M})}$~(Eq.(\ref{eq: FIB_Loss}));
    \STATE Compute graph-level information bottleneck loss $\hat{HSIC(\mathbf{S}',\mathbf{S})}$ (Eq.(28));
    \STATE Obtain the overall optimization loss at the first stage  $\mathcal{L}_1$~(Eq.(30));
    \STATE Update all parameters according to Eq.(\ref{eq: all_loss});
    \STATE Compute the optimization loss at the second stage $\mathcal{L}_{2}$~(Eq.(20));
    \STATE Update multimedia preference parameters $\Theta_2=[\mathbf{C}, \mathbf{Z}]$ according to Eq.(31);
    \ENDWHILE
\STATE Return $\Theta_1^{*}=[\mathbf{P^*}, \mathbf{Q^*}, \mathbf{C^*}, \mathbf{Z^*}]$.
\end{algorithmic}
\end{algorithm}

\section{Experiments}
We conduct extensive experiments to verify the effectiveness of our proposed \shortname~. Specifically, we aim to answer the following questions: (1) How does \shortname~perform on both effectiveness and robustness compared with other multimedia recommendation methods? (2) How do the key components boost the performance of \shortname~? (3) How sensitive is \shortname~under different hyper-parameter settings?

\begin{table}[t]
    \centering 
	\setlength{\belowcaptionskip}{5pt} %
	\caption{The statistics of three datasets.}\label{tab:statistics}
    \scalebox{0.95}{\begin{tabular}{c|c|c|c|c|c|c}
    \hline
    Datasets & Users & Items & Interactions & Density & V & T \\ \hline
    Clothing & 39,387 & 23,033 & 237,488 & 0.026\% & 4,096 & 1,024 \\ \hline
    Sports & 35,598 & 18,357 & 256,308 & 0.039\% & 4,096 & 1,024 \\ \hline
    Baby & 19,455 & 7,050 & 139,110 & 0.101\% & 4,096 & 1,024 \\ \hline
    \end{tabular}}
    \label{tab: datasets}
\end{table}

\subsection{Experimental Settings}
\subsubsection{Dataset Description}
We conduct empirical studies on three widely used Amazon multimedia recommendation datasets: Clothing, Sports, and Baby~\cite{mcauley2015image, zhang2021LATTICE}. Each dataset includes user-item interactions, item visual modality features, and item textual multimedia features. Following the existing multimedia recommendation works~\cite{mcauley2015image, zhang2021LATTICE}, we use the pre-trained multimedia features, which the visual features are extracted from the pre-trained VGG network~\cite{simonyan2014VGG}, and the textual features are extracted from the pre-trained Sentence-Bert~\cite{reimers2019sentence}, all features are released in~\cite{mcauley2015image, zhang2021LATTICE}. Detailed statistics of the used datasets are summarized in Table \ref{tab: datasets}.


\subsubsection{Baselines and Evaluation Metrics} 
\textbf{Baslines.} We compare our proposed method with SOTA methods, which can be divided into three groups: (1) Collaborative Filtering~(BPR-MF, LightGCN, SGL); (2) Multimedia-enhanced Recommendation Methods~(VBPR, MMGCN, GRCN, LATTICE); (3) Denoised Multimedia Recommendation Methods~(MICRO, InvRL, SGFD).
\begin{itemize}
    \item \textbf{BPR-MF}~\cite{UAI2009BPR}: is a classic collaborative filtering method based on matrix factorization. It employs pair-wise ranking optimization, which follows the basic that the observed interactions have higher scores compared with those unobserved.
    \item \textbf{LightGCN}~\cite{LightGCN}: simplifies graph neural collaborative filtering method, which removes the redundant non-linear activation and feature transformation.
    \item \textbf{SGL}~\cite{SGL} introduces the self-supervised graph learning technique to collaborative filtering, and designs three kinds of graph data augmentation strategies for contrastive learning.
    \item \textbf{VBPR}~\cite{he2016VBPR}: extents BPR with pre-trained item content features, VBPR learns user preference from both collaborative and content spaces.
    \item \textbf{MMGCN}~\cite{wei2019MMGCN}: constructs modality-specific graphs and refines modality-specific representation for users and items with GNNs.
    \item \textbf{GRCN}~\cite{wei2020GRCN}: refines the user-item interaction graph by identifying the false-positive interactions based on modality information.
    \item \textbf{LATTICE}~\cite{zhang2021LATTICE}: constructs the latent item-item structure based on multimodal features, then injects high-order item affinities into item representation.
    \item \textbf{InvRL}~\cite{du2022invariant}: proposes to alleviate the spurious correlations from the pre-trained multimedia features based on invariant representation learning.
    \item \textbf{MICOR}~\cite{MICOR}: designs a multimodal contrastive framework to alleviate the modality noise issue. It maximizes the agreement between the modality-aware representation and modality-fused representation.
    \item \textbf{SGFD}~\cite{SGFP}: utilizes semantic-guided feature distillation to extract denoised multimedia features to enhance multimodal recommendation.
\end{itemize}

\begin{table*}[th]
\centering
\caption{Overall comparisons of our proposed \shortname~ with different baselines in terms of Recall@20~(R@20), Precision@20~(P@20), NDCG@20~(N@20) on three datasets. The best performance is highlighted in \textbf{bold} and the second is highlighted by \underline{underlines}. Improvement indicates the relative improvement of our proposed \shortname~ compared to the best baseline in percentage.}
\label{tab: overall performance}
\scalebox{1.22}{
\begin{tabular}{l|ccc|ccc|ccc}
\hline 
 & \multicolumn{3}{c}{Clothing}       
 & \multicolumn{3}{c}{Sports}
 & \multicolumn{3}{c}{Baby}\\ \cline{2-10}
\multirow{-2}{*}{Models} & R@20 & P@20 & N@20 & R@20 & P@20 & N@20 & R@20 & P@20& N@20\\ \hline
BPR-MF & 0.0267 & 0.0014 & 0.0122 & 0.0658& 0.0035 & 0.0308 & 0.0612 & 0.0033 & 0.0264\\ 
LightGCN & 0.0547 & 0.0028 & 0.0245 & 0.0846 & 0.0045 & 0.0393 & 0.0756 & 0.0040 & 0.0336\\ 
SGL & 0.0598 & 0.0030 & 0.0268 & 0.0905 & 0.0047 & 0.0412 & 0.0745 & 0.0040 & 0.0328\\ \hline 

VBPR & 0.0464 & 0.0023 & 0.0202 & 0.0678 & 0.0036 & 0.0315 & 0.0677 & 0.0036 & 0.0299 \\ 
MMGCN & 0.0501 & 0.0024 & 0.0221 & 0.0638 & 0.0034 & 0.0279 & 0.0640 & 0.0032 & 0.0284\\ 
GRCN & 0.0631 & 0.0032 & 0.0276 & 0.0833 & 0.0044 & 0.0377 & 0.0754 & 0.0040 & 0.0336\\ 
LATTICE & 0.0710 & 0.0036 & 0.0316 & 0.0915 & 0.0048 & 0.0424 & 0.0829 & 0.0044 & 0.0368 \\ \hline 
InvRL & 0.0729 & 0.0037 & 0.0326 & 0.0957 & 0.0050 & 0.0431 & 0.0857 & 0.0046 & 0.0391  \\ 
MICOR & \underline{0.0782} & \underline{0.0040} & \underline{0.0351} & \underline{0.0988} & \underline{0.0052} & \underline{0.0457} & \underline{0.0892} & \underline{0.0047} & \underline{0.0402} \\ 
SGFD & 0.0746 & 0.0038 & 0.0334 & 0.0971 & 0.0051 & 0.0443 & 0.0853 & 0.0046 & 0.0386\\ \hline \hline

\textbf{\shortname} & \textbf{0.0891} & \textbf{0.0045} & \textbf{0.0401} &  \textbf{0.1053} & \textbf{0.0056} & \textbf{0.0496} & \textbf{0.0970} & \textbf{0.0051} & \textbf{0.0438} \\ 
\textbf{Improve.} & \textbf{13.94\%} & \textbf{12.50\%} & \textbf{14.25\%} & \textbf{6.58\%}& \textbf{7.69\%} & \textbf{8.53\%} & \textbf{8.74\%} & \textbf{8.51\%} & \textbf{8.96\%} \\ \hline
\end{tabular}}
\end{table*}

\textbf{Evaluation Metrics.}
As we focus on the ranking task, we employ three widely used evaluation metrics: Recall@N, Precision@N, and NDCG@N for the Top-N ranking item list. For all metrics, we use full full-ranking strategy that selects all non-interacted items as candidates. All metrics are reported with average values of 10 times repeated experiments.

\subsubsection{Implement Details}
We implement our method in Tensorflow-GPU~\footnote{https://www.tensorflow.org/} with TITAN RTX~(24G). We initialize model embeddings with a Gaussian distribution with a mean value of 0 and a standard variance of 0.01. We fix the latent embedding size to 64, and each modality feature also be projected into 64 dimension space. For model training, we use Adam optimizer with a learning rate of 0.0005. We adopt the batch training strategy with a batch size of 1024, and the training data consists of triplets. For all baselines, we refer to the parameters reported in the original paper and carefully fine-turn them with grid-search. Besides, we employ an early-stop strategy for model training, that the model will stop training if Recall@20 on the validation doesn't increase for 10 epochs to avoid over-fitting. 


\subsection{Performance Comparison}
We report the recommendation performances of our \shortname and other baselines over three datasets. As shown in Table \ref{tab: overall performance}, we have the following observations:
\begin{itemize}
    \item Compared with CF methods, multimedia-enhanced recommendation methods have better performances whether representation enhancement~(VBPR over BPR) and item-item correlation formulation~(LATTICE over LightGCN). Besides, standing on the better model architecture, i.e., MF to GCN, multimedia content consistently improves collaborative filtering methods. These demonstrate the effectiveness of leveraging multimedia information to boost recommendation tasks.
    
    \item Although multimedia content significantly boosts recommenders, learning denoised multimedia representation is still necessary. We select the competing multimedia-enhanced recommendation models~(LATTICE) as the backbone, and then deploy three current multimedia denoising methods~(InvRL, MICOR, SGFD). Compared with LATTICE, all multimedia denoising methods show better performances over three datasets. This indicates that directly using the pre-trained multimedia features is sub-optimal, there exist semantic gaps between domain-specific pre-trained multimedia features and recommendation scenarios.

    \item Our proposed \shortname~ consistently outperforms all baselines over all datasets, verifying the effectiveness of the proposed method. Firstly, compared with CF-based methods, \shortname~ leverages multimedia features to enhance recommendation from both representation enhancement and item-item correlation formulation. Furthermore, compared with Multimedia-enhanced recommendation methods, \shortname~ removes the irrelevant multimedia features and then narrows the gap between pre-trained features and recommendation tasks. Finally, compared with the strongest baseline, i.e., MICOR, \shortname~ achieves $w.r.t$ NDCG@20 by 14.25\%, 8.53\%, and 8.96\% relative improvements on the Clothing, Sports, and Baby dataset, respectively. This demonstrates the superiority of removing task-irrelevant multimedia features via the Information Bottleneck principle over other multimedia-denoising methods. Given feature-level and graph-level Information Bottleneck learning, our \shortname~ can significantly improve multimedia-based recommendation under various settings.
\end{itemize}


\begin{table*}[t]
\small
\centering 
\caption{Performances of our proposed \shortname~ under different backbones.Among them, VBPR and VLightGCN leverage multimedia features to enhance representation, so \shortname~ only performs Feature-level IB~(FIB) learning on them. LATTICE leverages multimedia features to learn the item-item correlation, so \shortname~ only performs Graph-level IB~(GIB) learning on it. VLATTICE is the extension version of LATTICE that combines representation enhancement, to this end, \shortname~ joint FIB and GIB learning to remove the irrelevant features to empower recommendation. }
\label{tab: backbones}
\scalebox{1.1}{
\begin{tabular}{l|c|c|c|c|c|c|c|c|c}
\hline 
 & \multicolumn{3}{c}{Clothing}       
 & \multicolumn{3}{c}{Sports}
 & \multicolumn{3}{c}{Baby}\\ \cline{2-10}
\multirow{-2}{*}{Models} & R@20 & P@20 & N@20 & R@20 & P@20 & N@20 & R@20 & P@20& N@20\\ \hline
VBPR & 0.0462 & 0.0023 & 0.0202 & 0.0678 & 0.0036 & 0.0315& 0.0677 & 0.0036 & 0.0299 \\ 
\rowcolor{gray!25} \shortname & 0.0558 & 0.0028 & 0.0248 & 0.0773 & 0.0041 & 0.0358 & 0.0754 & 0.0040 & 0.0336
 \\ 
$\triangle$ Imp. & 20.78\% & 21.74\% & 22.77\% & 14.01\% & 13.89\% & 13.65\% & 11.37\% & 11.11\% & 12.37\% \\ \hline \hline
VLightGCN & 0.0627 & 0.0032 & 0.0282 & 0.0946 & 0.0050 & 0.0442 & 0.0870 & 0.0046 & 0.0386\\ 
\rowcolor{gray!25} \shortname & 0.0686 & 0.0035 & 0.0307 & 0.0990 & 0.0052 & 0.0461 & 0.0912 & 0.0048 & 0.0403 \\ 
$\triangle$ Imp. & 9.41\% & 9.38\% & 8.87\% & 4.65\% & 4.00\% & 3.95\% & 4.30\% & 4.35\% & 4.40\% \\ \hline \hline

LATTICE & 0.0710 & 0.0036 & 0.0316 & 0.0915 & 0.0048 & 0.0424 & 0.0829 & 0.0044 & 0.0368 \\
\rowcolor{gray!25} \shortname~ & 0.0788 & 0.0040 & 0.0349 & 0.0988 & 0.0052 & 0.0451 & 0.0907 & 0.0048 & 0.0403 \\
$\triangle$ Imp. & 10.99\% & 11.11\% & 10.44\% & 7.98\% & 8.33\% & 6.37\% & 9.41\% & 9.09\% & 9.51\%\\ \hline

VLATTICE & 0.0699 & 0.0035 & 0.0316 & 0.0972 & 0.0051 & 0.0449 & 0.0878 & 0.0046 & 0.0390\\ 
\rowcolor{gray!25} \shortname & 0.0891 & 0.0045 & 0.0401 & 0.1053 & 0.0056 & 0.0496 &  0.0970 & 0.0051 & 0.0438 \\ 
$\triangle$ Imp. & 27.47\% & 28.57\% & 26.90\% 
 & 8.33\% & 9.80\% & 10.47\% & 10.48\% & 10.87\% & 12.31\%
\\ \hline
\end{tabular}}
\end{table*}

\subsection{Investigation of \shortname~}
\subsubsection{Generality with Various Backbones}
Our proposed \shortname~ is a general multimedia representation learning module, which can easily equipped with various multimedia-enhanced recommendation methods. In this section, we conduct experiments to demonstrate the generality of the proposed \shortname~ to improve various multimedia recommendation backbones.  We select four representative recommendation backbones to conduct the generality experiments: VBPR, VlightGCN, LATTICE, and VLATTICE. Among them, VBPR and VlightGCN are extensions of BPR and LightGCN, which concatenate content embedding to enhance item representation, so
\shortname~ only performs Feature-level IB~(FIB) learning on them.
Instead of representation enhancement, LATTICE leverages multimedia features to learn item-item correlation, so \shortname~ only performs Graph-level IB~(GIB) learning on it. VLATTICE is the extension version of LATTICE that combines representation enhancement and graph structure learning. To this end, \shortname~ joint FIB and GIB learning to remove the impact of irrelevant features from representation and structure perspectives. 

From Table \ref{tab: backbones}, we can observe that our proposed \shortname~ significantly improves all backbones on three datasets, which strongly demonstrates that \shortname~ has well generality combined with SOTA multimedia-based recommendation methods. For representation enhancement multimedia recommenders~(VBPR, VLightGCN), \shortname~ uses FIB learning to remove the impact of irrelevant features and panned out. Especially on the VBPR backbone, \shortname~ achieves impressive improvements on three datasets, i.e., over 20\% gains on the Clothing dataset. For structure learning based multimedia recommender~(LATTICE), \shortname~ uses GIB learning to remove the impact of irrelevant features and also achieves success. More stable item-item structures make considerable improvements on three datasets, i.e., over 10\% gains on the Clothing dataset. Furthermore, compared with VLATTICE, \shortname~ shows great improvements over three datasets, which demonstrate the effectiveness of combining FIB and GIB learning to remove irrelevant multimedia features for recommendation. The above observations verify the effectiveness and generality of our proposed \shortname~, which can significantly empower current multimedia recommendation methods.

\begin{table*}[t]
    \centering 
    \caption{Ablation study of \shortname. 
    \shortname-w/o IB denotes without any Information Bottleneck learning, \shortname-w/o FIB denotes without Feature-level Information Bottleneck learning, \shortname-w/o GIB denotes without Graph-level Information Bottleneck learning.}
    \label{tab: ablation study}
    \scalebox{1.22}{\begin{tabular}{l|ccc|ccc|ccc}
    \hline 
    & \multicolumn{3}{c}{Clothing}       
    & \multicolumn{3}{c}{Sports}
    & \multicolumn{3}{c}{Baby}\\ \cline{2-10}
    \multirow{-2}{*}{Models} & R@20 & P@20 & N@20 & R@20 & P@20 & N@20 & R@20 & P@20& N@20\\ \hline
    \shortname-$w/o$ IB & 0.0699 & 0.0035 & 0.0316 & 0.0972 & 0.0051 & 0.0449 & 0.0878 & 0.0046 & 0.0390 \\ \hline
    \shortname-$w/o$ FIB & 0.0741 & 0.0038 & 0.0331 & 0.1014 & 0.0054  & 0.0471 & 0.0928 & 0.0049 & 0.0413 \\ \hline
    \shortname-$w/o$ GIB & 0.0759 & 0.0038 & 0.0336 & 0.1038 & 0.0055 & 0.0486 & 0.0935 & 0.0049 & 0.0425
    \\ \hline
    \textbf{\shortname} & \textbf{0.0891} & \textbf{0.0045} & \textbf{0.0401} & \textbf{0.1053} & \textbf{0.0056} & \textbf{0.0496} & \textbf{0.0970} & \textbf{0.0051} & \textbf{0.0438} \\ \hline
    \end{tabular}}
\end{table*}

\subsubsection{Ablation Study}
To exploit the contribution of each component of \shortname~ with VLATTICE backbone, we conduct ablation studies on three datasets. 
As shown in Table \ref{tab: ablation study}, we compare \shortname~ and its corresponding variants on Top-20 recommendation performance. 
Specifically, \shortname-$w/o$ FIB and \shortname-$w/o$ GIB denote removing the FIB and GIB learning modules, respectively. Besides, \shortname-$w/o$ IB denotes that removing all IB learning modules, then \shortname~ degenerates to VLATTICE. From Table \ref{tab: ablation study}, we have the following observations. First, \shortname-$w/o$ IB shows the worst performance due to it discarding any Information Bottleneck learning, the generic multimedia features introduce irrelevant information to hinder preference learning. Compared with \shortname-$w/o$ IB, \shortname-$w/o$ FIB and \shortname-$w/o$ GIB all have better performances. It indicates that both Feature-level and Graph-level Information Bottleneck learning reduce irrelevant features to enhance multimedia recommendation. Finally, combining Feature-level and Graph-level Information Bottleneck learning, \shortname~ achieves the best performances on three datasets, which verifies the effectiveness of removing the impact of irrelevant features from feature and structure perspectives.

\begin{figure*} [th]
  \begin{center}
      \subfigure[Clothing]{
      \includegraphics[width=52mm]{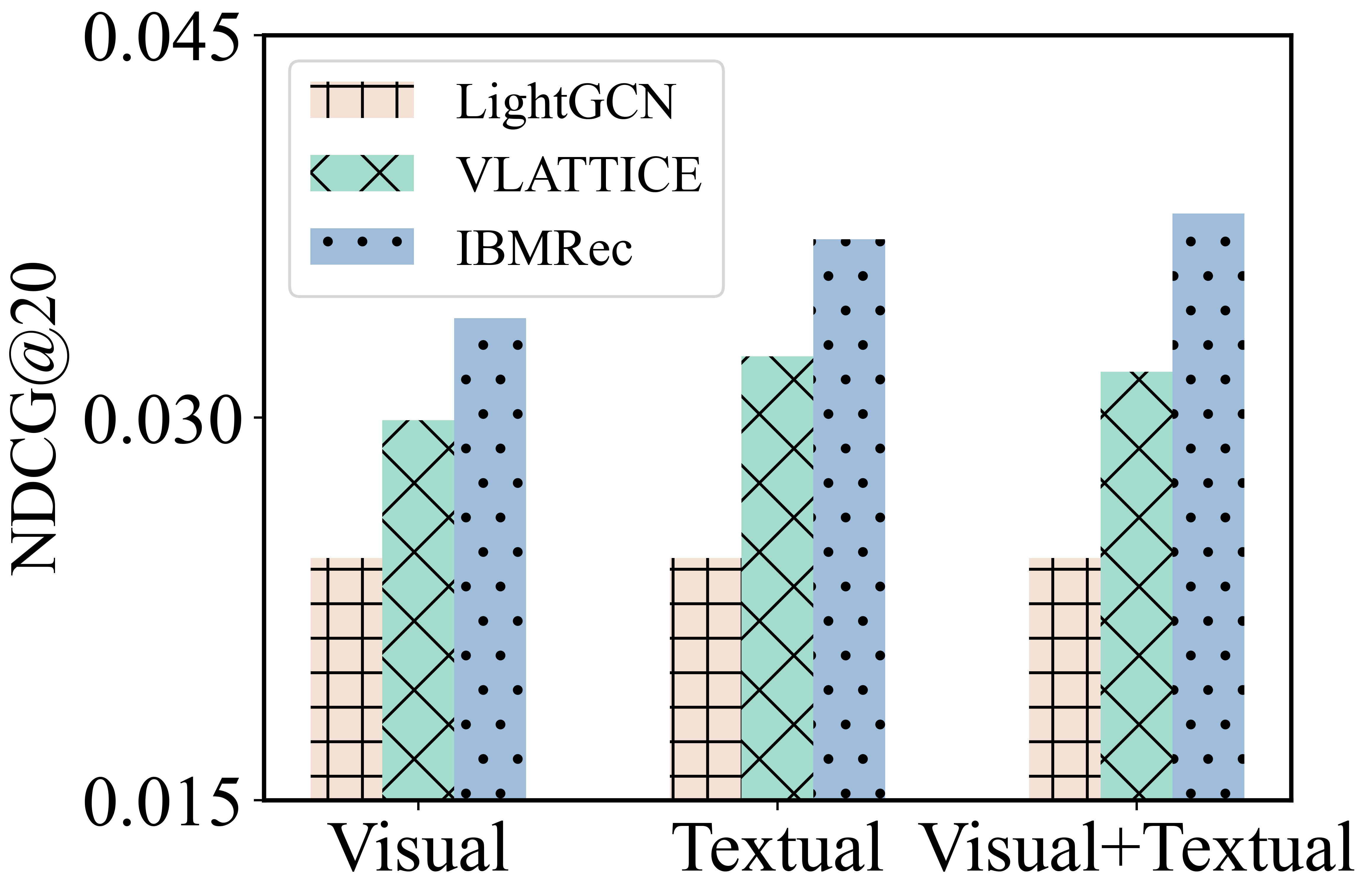}}
      \subfigure[Sports]{
      \includegraphics[width=52mm]{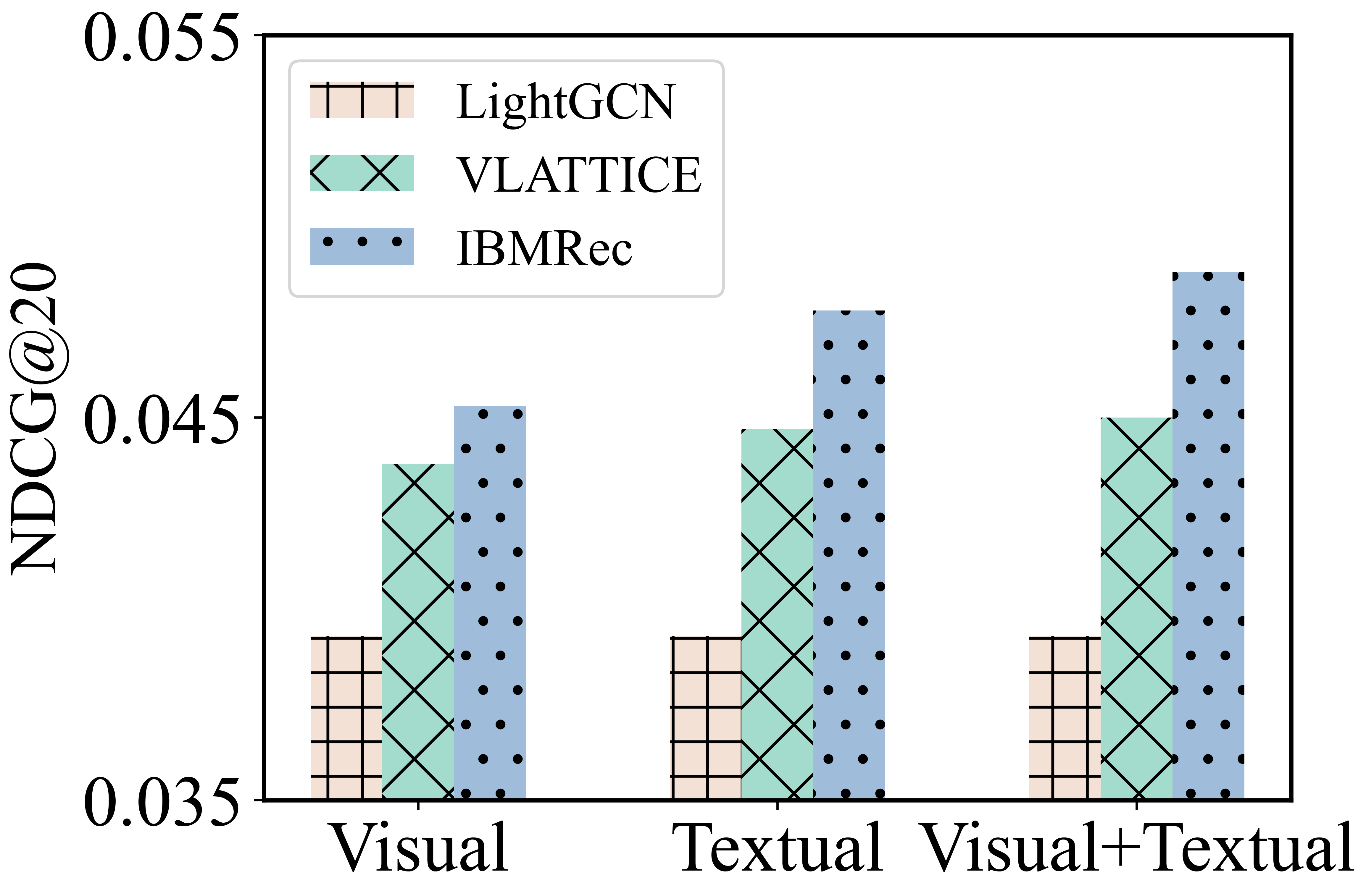}}
      \subfigure[Baby]{
      \includegraphics[width=52mm]{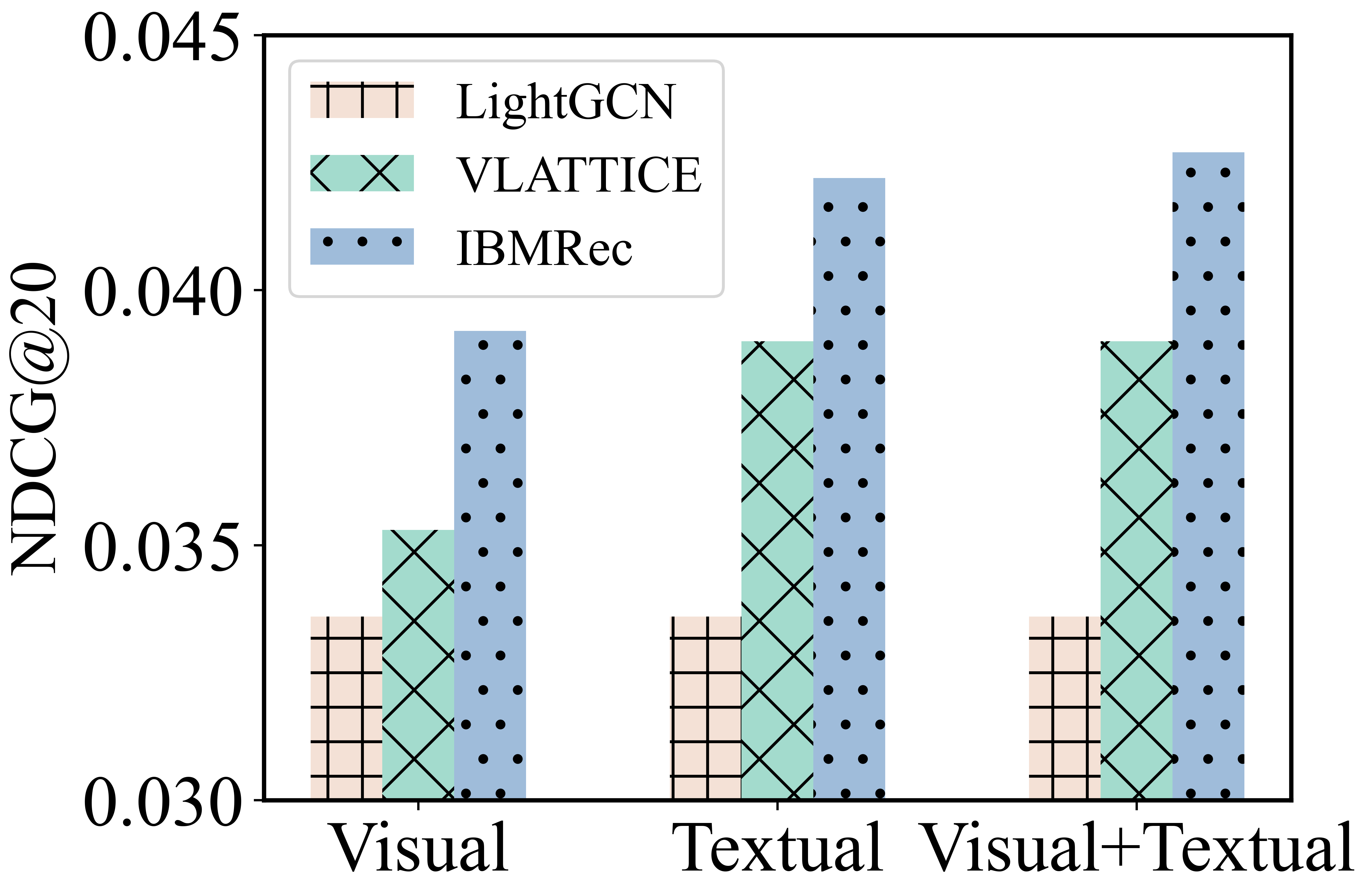}}
  \end{center}
  \caption{\small{Performance comparisons under different modalities.}} 
  \label{fig: modality}
\end{figure*}

\subsection{Detailed Analysis of \shortname}
\subsubsection{Effect of Different Modalities}
Here we investigate the effect of multi-modality information for recommendation. As illustrated in Fig. \ref{fig: modality}, we compare three recommendation models under different modalities. Specifically, LightGCN is the compared CF backbone, and VLATTICE is the simplified version of \shortname~ without IB learning. From Fig. \ref{fig: modality}, we have the following observations:
\begin{itemize}
    \item Compared to LightGCN, VLATTICE consistently demonstrates superior performance under both visual and textual modalities. This observation substantiates that each modality feature contributes significantly to recommendation performance. Furthermore, in comparison between VLATTICE~(with the visual feature) and VLATTICE~(with the textual feature) across three datasets, the latter consistently exhibits better performance. This suggests that the textual modality imparts additional information that enhances the overall quality of recommendations.
    \item In contrast to VLATTICE, \shortname~ exhibits noteworthy improvements under each modality. This underscores the effectiveness of our designed IB learning modules in adeptly eliminating irrelevant multimedia features. Moreover, it highlights the robustness of these modules, demonstrating their insensitivity to variations across different modalities.
    \item In our comparison of single and multiple modalities, VLATTICE's attempt to combine visual and textual features didn't yield significant improvements, especially showing poorer performance on the Clothing dataset. This suggests a limitation in VLATTICE's ability to integrate multi-modal information effectively for recommendations. Fortunately, our proposed \shortname~ addresses this challenge seamlessly. Experimental results demonstrate that, for \shortname~, the inclusion of more modalities leads to better performances. This can be attributed to our approach of removing task-irrelevant features for each modality, allowing all modalities to collaborate effectively in enhancing recommendation tasks.
\end{itemize}


\begin{figure*} [th]
  \begin{center}
      \subfigure[Clothing]{
      \includegraphics[width=52mm]{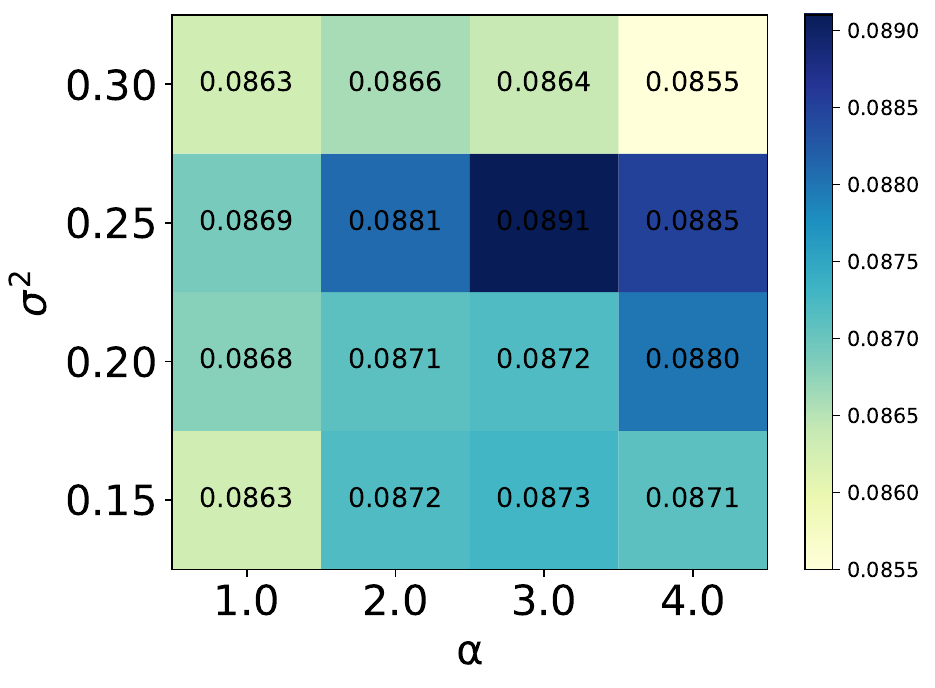}}
      \subfigure[Sports]{
      \includegraphics[width=52mm]{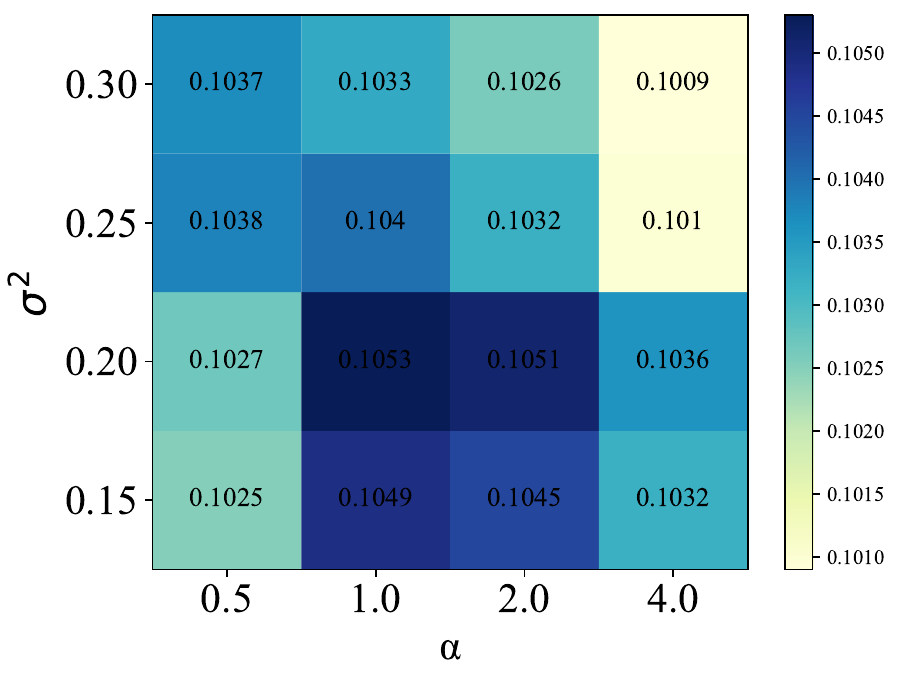}}
      \subfigure[Baby]{
      \includegraphics[width=52mm]{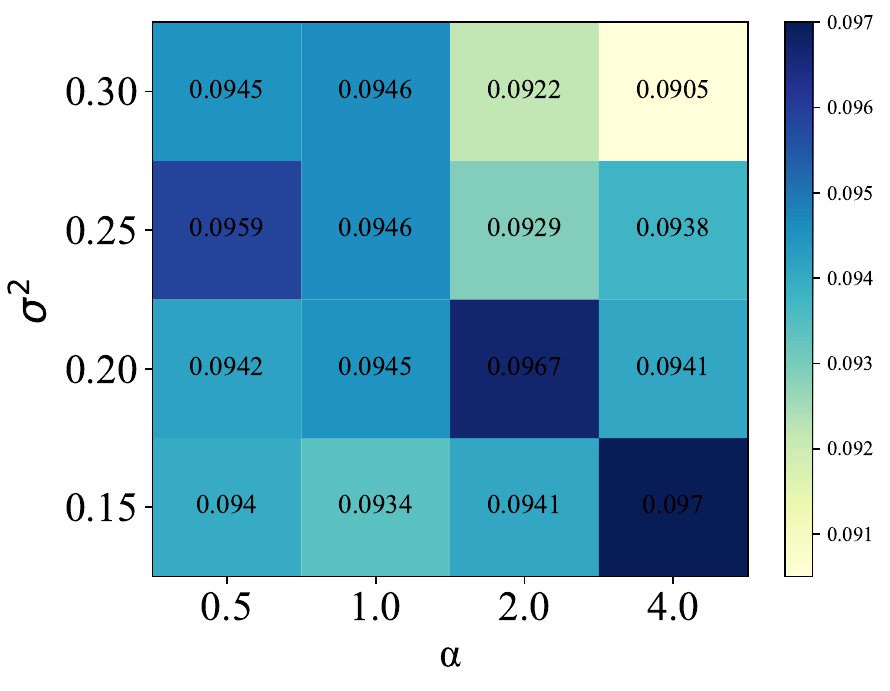}}
  \end{center}
  \vspace{-0.3cm}
  \caption{\small{Impact of different FIB loss parameters $\alpha$ and $\sigma^2$.}} 
  \label{fig: hyper-parameters}
\end{figure*}

\begin{figure*} [th]
  \begin{center}
      \subfigure[Clothing]{
      \includegraphics[width=52mm]{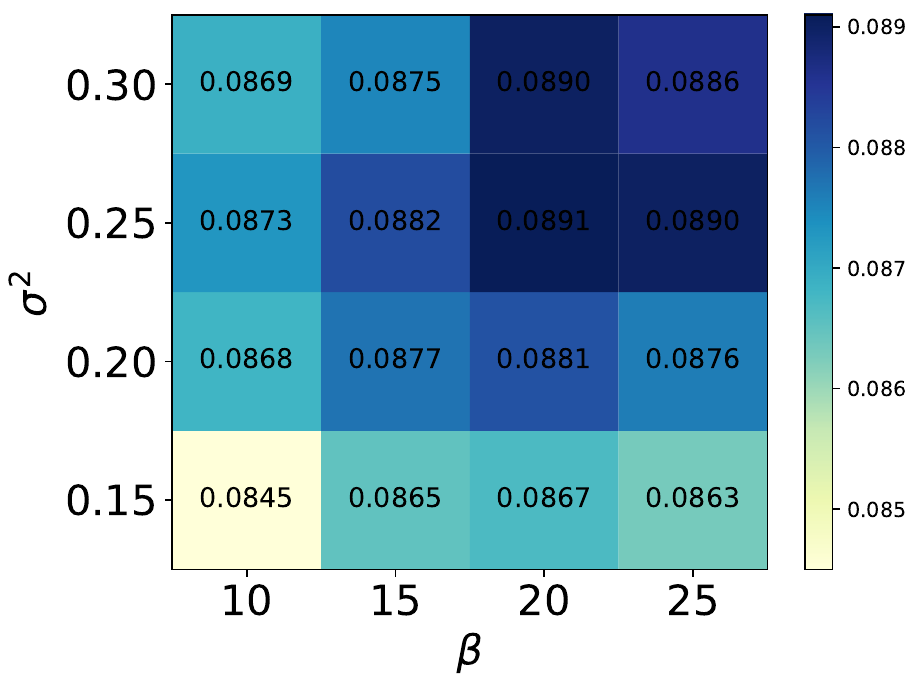}}
      \subfigure[Sports]{
      \includegraphics[width=52mm]{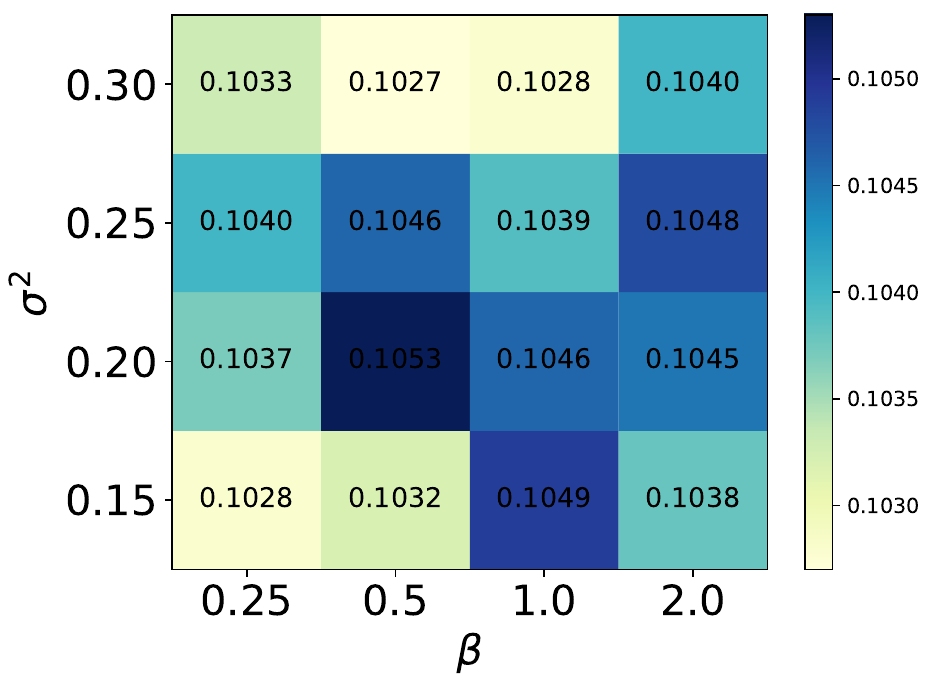}}
      \subfigure[Baby]{
      \includegraphics[width=52mm]{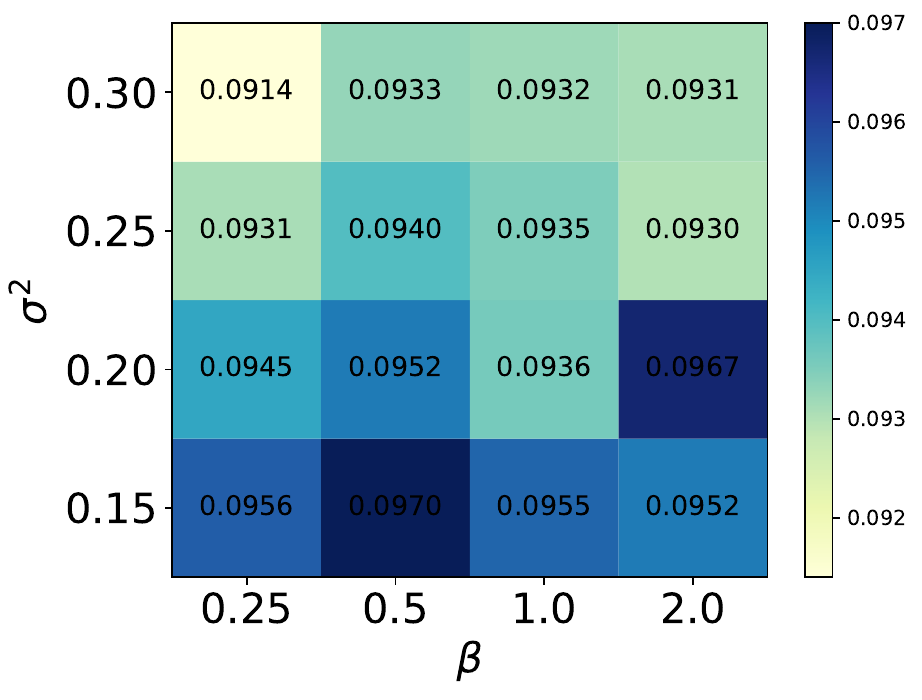}}
  \end{center}
  \vspace{-0.3cm}
  \caption{\small{Impact of different GIB loss parameters $\beta$ and $\sigma^2$.}} 
  \label{fig: hyper-parameters}
\end{figure*}

\subsubsection{Parameter Sensitivities}
In this section, we delve into the impact of hyperparameters in \shortname~, specifically examining the FIB loss coefficient $\alpha$, GIB loss coefficient $\beta$, and RBF parameter $\sigma^2$. These parameters determine the trade-off between removing irrelevant features and using multimedia information. 

We analyze the impact of different scale FIB objectives $(\alpha, \sigma^2)$. Throughout our experiments, we explored $(\alpha, \sigma^2)$ across a range of values, to assess their influence on the model's performances.
As shown in Fig. \ref{fig: hyper-parameters}, \shortname~reaches the best performance when $(\alpha=3.0, \sigma^2=0.25)$, $(\alpha=1.0, \sigma^2=0.20)$, and $(\alpha=1.0, \sigma^2=0.15)$ on the Clothing, Sports, and Baby datasets, respectively.
Additionally, we investigate the influences of different scale GIB objectives $(\beta, \sigma^2)$. 
As shown in Fig. \ref{fig: hyper-parameters}, \shortname~reaches the best performance when $(\beta=20, \sigma^2=0.25)$, $(\alpha=0.5, \sigma^2=0.25)$, and $(\alpha=0.5, \sigma^2=0.15)$ on the Clothing, Sports, and Baby datasets, respectively.
This comprehensive exploration allows us to gain insights into how different parameter configurations affect the overall model performance. 


\section{Conclusion}
In this work, we investigate the irrelevant feature problem in multimedia recommendations, and propose a novel \shortname~via the Information Bottleneck principle. Specifically, \shortname~consists of two elaborate Information Bottleneck learning modules: Feature-level IB learning~(FIB) and
Graph-level IB learning~(GIB). FIB first designs a decomposed mutual information maximization, then introduces the the HSIC bottleneck to reduce the dependency between multimedia representation and its original features, which can effectively remove irrelevant features for recommendation. Considering the noise-amplifying problem with an unstable item-item graph, we further propose the GIB learning module, which reconstructs a preference-guided graph to better exploit item affinity into preference learning. Besides, our proposed \shortname~is a general multimedia denoising module, which can be flexiblely coupled with multimedia recommenders. Experiments conducted on three public datasets, and empirical studies verify the effectiveness of our proposed \shortname~, showcasing high performance, and applicability to various multimedia recommenders. 



\begin{small}
\bibliographystyle{abbrv}
\bibliography{IBRec}
\end{small}

\begin{IEEEbiography}[{\includegraphics[width=1in,height=1.25in,clip,keepaspectratio]{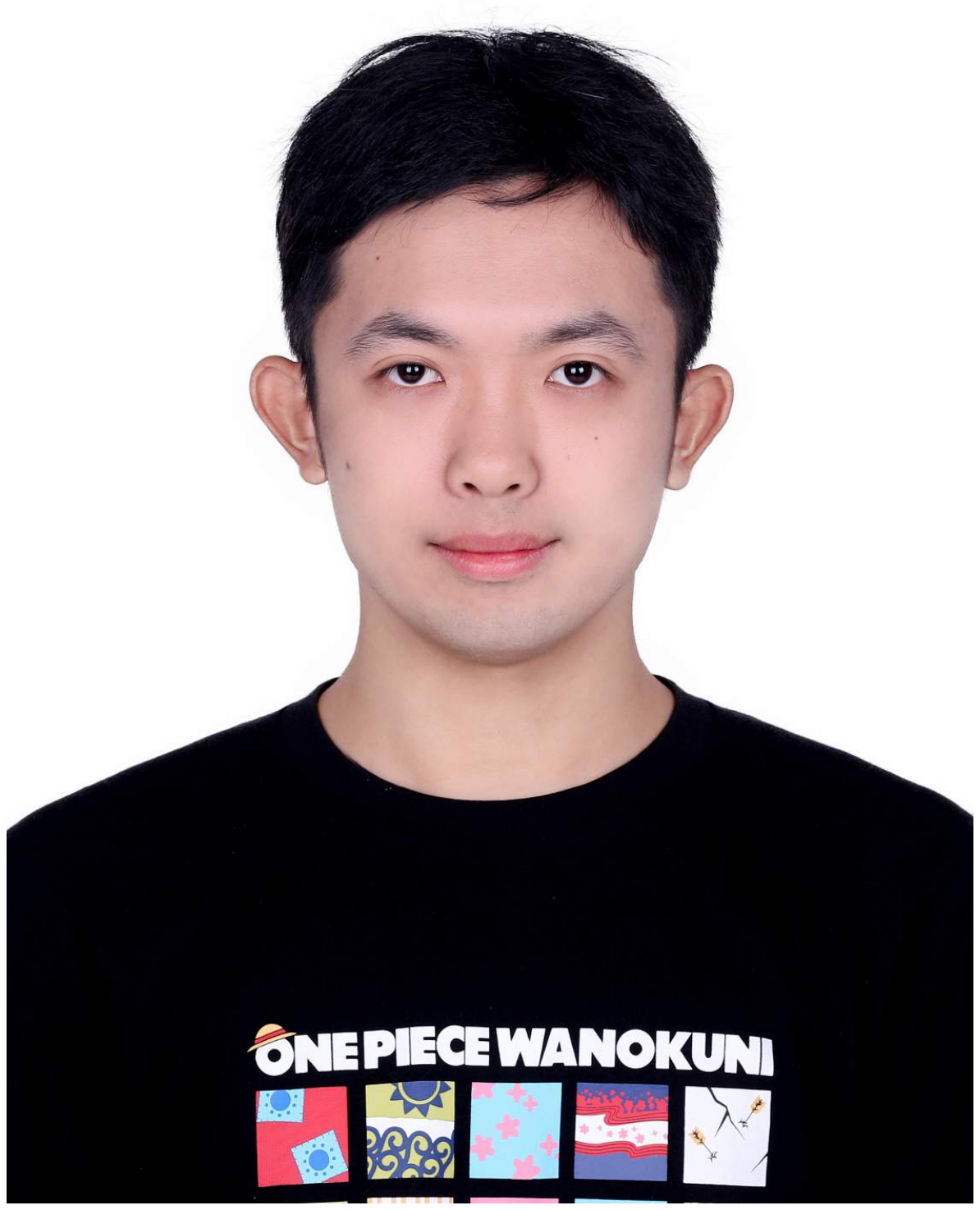}}]
{Yonghui Yang} is currently pursuing a Ph.D. degree at Hefei University of Technology, China. He obtained his master's degree from the same university in 2021. He has published several papers in leading journals and conferences, such as IEEE TKDE, IEEE TBD, KDD, SIGIR, ACM Multimedia, and IJCAI. His research focuses on graph learning and recommender systems.
\end{IEEEbiography}

\begin{IEEEbiography}[{\includegraphics[width=1in,height=1.25in,clip,keepaspectratio]{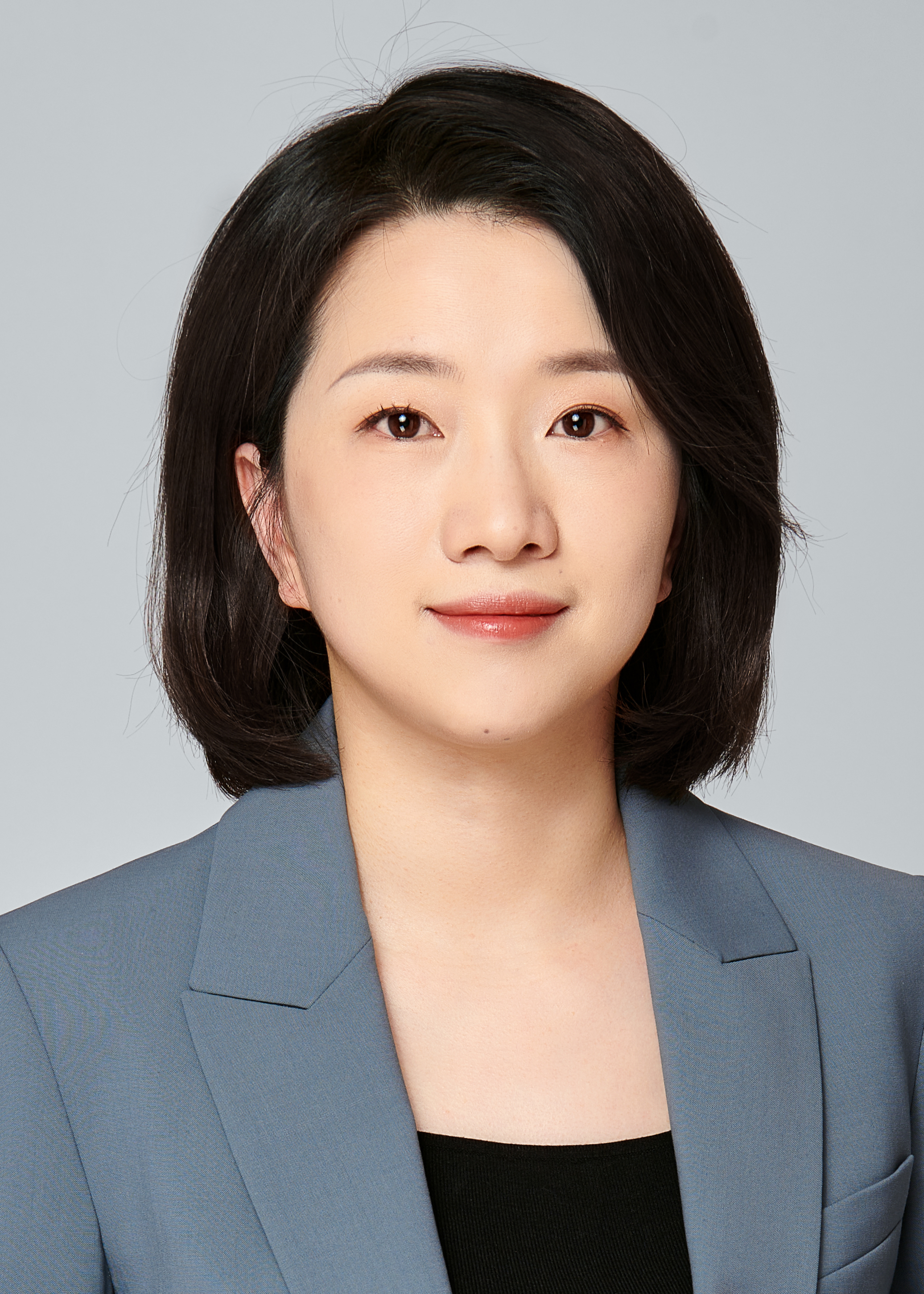}}]
{Le Wu} is currently a professor at the Hefei University of Technology (HFUT), China. She received her Ph.D. degree from the University of Science and Technology of China (USTC). Her research interests are data mining, recommender systems, and social network analysis. She has published more than 50 papers in referred journals and conferences, such as IEEE TKDE, NeurIPS, KDD, SIGIR, WWW, and AAAI. Dr. Le Wu is the recipient of the Best of SDM 2015 Award, and the Distinguished Dissertation Award from the China Association for Artificial Intelligence (CAAI) 2017.
\end{IEEEbiography}

\begin{IEEEbiography}[{\includegraphics[width=1in,height=1.25in,clip,keepaspectratio]{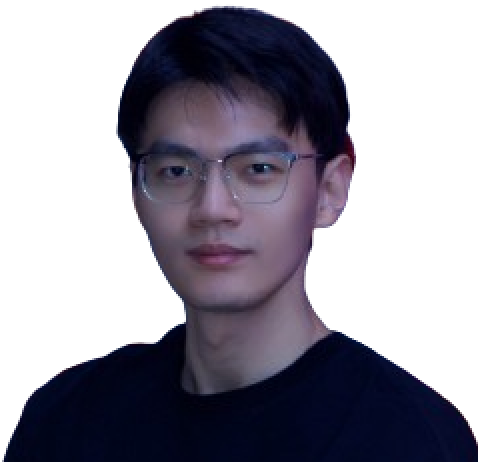}}]
{Zhuangzhuang He} is currently pursuing a master's degree at Hefei University of Technology, China. He obtained his bachelor's degree from the Anhui Agricultural University in 2022. He has published papers in leading conferences, including KDD, SIGIR. His research focuses on out-of-distribution generalization and recommender systems. 
\end{IEEEbiography}

\begin{IEEEbiography}
[{\includegraphics[width=1in,height=1.25in,clip,keepaspectratio]{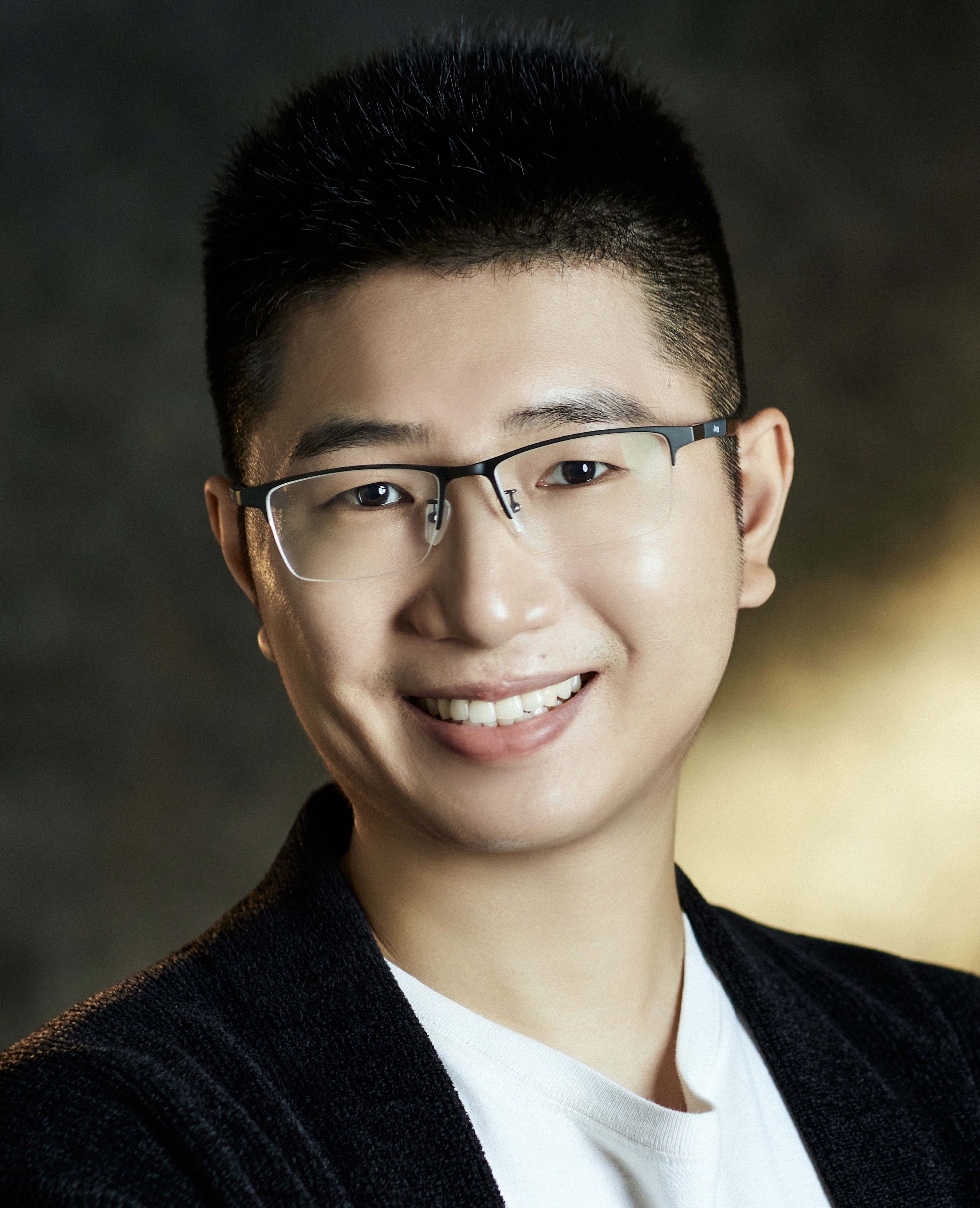}}]
{Zhengwei Wu} is presently employed at ByteDance in Hangzhou, China. He has published several papers in leading conferences, including NeurIPS, KDD, IJCAI, SIGIR. His current research interests include graph neural networks and recommender systems.
\end{IEEEbiography}

\begin{IEEEbiography}[{\includegraphics[width=1.0in,height=1.25in,clip,keepaspectratio]{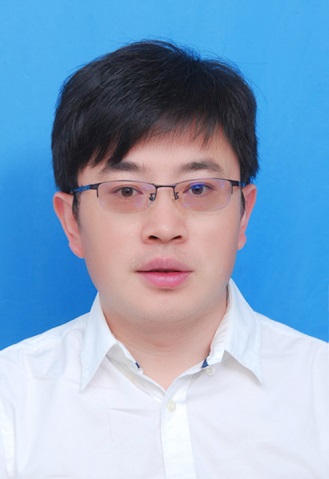}}]
{Richang Hong} is currently a professor at Hefei University of Technology. He received the Ph.D. degree from USTC, in 2008. He has published more than 100 publications in the areas of his research interests, which include multimedia question answering, video content analysis, and pattern recognition. He is a member of the Association for Computing Machinery. He was a recipient of the Best Paper award in the ACM Multimedia 2010.
\end{IEEEbiography}

\begin{IEEEbiography}[{\includegraphics[width=1.0in,height=1.25in,clip,keepaspectratio]{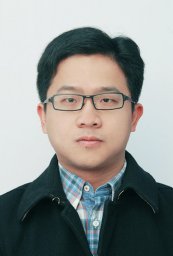}}]
{Meng Wang} (Fellow, IEEE) is a professor at the Hefei University of Technology, China. He received his B.E. degree and Ph.D. degree in the Special
Class for the Gifted Young and the Department of Electronic
Engineering and Information Science from the University of Science and
Technology of China (USTC), Hefei, China, in 2003 and 2008,
respectively. His current research interests include multimedia
content analysis, computer vision, and pattern recognition. He has
authored more than 200 book chapters, journal and conference papers in
these areas. He is the recipient of the ACM SIGMM Rising Star Award 2014.
He is an associate editor of IEEE Transactions on Knowledge and Data
Engineering (IEEE TKDE), IEEE Transactions on Circuits and Systems
for Video Technology (IEEE TCSVT), IEEE Transactions on Multimedia (IEEE TMM), and IEEE Transactions on Neural Networks and Learning Systems (IEEE TNNLS).
\end{IEEEbiography}
\end{document}